\documentclass[english]{article}
\usepackage[T1]{fontenc}
\usepackage[latin9]{inputenc}
\usepackage{geometry}
\usepackage{float}
\usepackage{amsmath}
\usepackage{amsbsy}
\usepackage{amstext}
\usepackage{amssymb}
\usepackage{graphicx}
\usepackage{esint}
\usepackage{comment}
\usepackage{datetime}
\usepackage{xcolor}
\usepackage{dirtytalk}
\usepackage{changepage}
\usepackage{enumitem}
\usepackage[round]{natbib}

\providecommand{\tabularnewline}{\\}

\usepackage{float}
\floatstyle{plaintop}
\restylefloat{table}

\usepackage[tableposition=top]{caption}

\makeatother

\usepackage{babel}

\begin{document}

\begin{center} \LARGE\textbf{Using a probabilistic approach to derive a two-phase model of flow-induced cell migration}\\[5mm] \normalsize Y. Ben-Ami$^1$, J.M. Pitt-Francis$^2$, P.K. Maini$^1$, H.M. Byrne$^{1,3}$\\[3mm] \end{center} 
\begin{enumerate}[label=\arabic*]
\small 
\item Wolfson Centre for Mathematical Biology, Mathematical Institute, University of Oxford, Oxford, United Kingdom
\item Department of Computer Science, University of Oxford, Oxford, United Kingdom \item Ludwig Institute for Cancer Research, University of Oxford, Oxford, United Kingdom
\end{enumerate}
\begin{center} \footnotesize \today  \end{center} \normalsize

\smallskip{}

\begin{abstract}
Interstitial fluid flow is 
a feature of many solid tumours. 
\textit{In vitro} experiments have shown that such fluid flow can
direct tumour cell movement upstream or downstream
depending on the balance between the competing mechanisms of tensotaxis (cell migration up stress gradients) and autologous chemotaxis (downstream cell movement in response to flow-induced gradients of self-secreted chemoattractants).

In this work we develop a probabilistic-continuum, two-phase model for cell migration in response to interstitial flow. We use a kinetic description for the cell-velocity probability density function, and model the flow-dependent mechanical and chemical stimuli as forcing terms which bias cell migration upstream and downstream. 
Using velocity-space averaging, we reformulate the model as a system of continuum equations for the spatio-temporal evolution of the cell volume fraction and flux, in response to forcing terms which depend on the local direction and magnitude of the mechanochemical cues. 

We specialise our model to describe a one-dimensional cell layer subject to fluid flow. Using a combination of numerical simulations and asymptotic analysis, we delineate the parameter regime where transitions from downstream to upstream cell migration occur.
As has been observed experimentally, 
the model predicts downstream-oriented, chemotactic migration at low cell volume fractions, and upstream-oriented, tensotactic migration at larger volume fractions. We show that the locus of the critical volume fraction, at which the system transitions from downstream to upstream migration, is dominated by the
ratio of the rate of chemokine secretion and advection.
Our model also predicts that, because the tensotactic stimulus depends strongly on the cell volume fraction, upstream, tensotaxis-dominated migration occurs only transiently when the cells are initially seeded, and transitions to downstream, chemotaxis-dominated migration occur at later times due to the dispersive effect of cell diffusion.

\end{abstract}

\section{Introduction}

Cells can sense a variety of chemical and mechanical cues 
which may bias their movement. In healthy tissues, cells  
migrate in response to multiple environmental cues; examples include
morphogenesis, wound healing and the stimulation of an immune response to 
infection \citep{sengupta2021principles}. At the same time, many diseases are characterised by excessive (or insufficient) directed cell migration; examples include tumour invasion and metastasis to adjacent tissues \citep{roussos2011chemotaxis, shields2007autologous}, and impaired wound healing caused by diabetes \citep{falanga2005wound}.

Fluid flow has been found to promote tumour cell migration in several different ways \citep{shields2007autologous,polacheck2011interstitial,polacheck2014mechanotransduction,lee2020integrated}.
Interstitial fluid flow in solid tumours is known to be higher 
than in healthy tissues due to growth-induced increases in interstitial
pressure and leaky blood vessels. Consequently, interstitial
flow has been suggested as a contributor to cell migration
and metastasis \citep{heldin2004high,follain2020fluids}.

{\it In vitro} experiments \cite{polacheck2011interstitial} have shown that fluid flow may impact the directed movement of cells in several different ways. On the one hand, extracellular fluid flow increases
the pressure on the upstream part of the cell and, consequently,
the cell increases the adhesion forces it exerts on the extracellular matrix (ECM) in this region.
In turn, the localized tension at the front of the cell leads
to actin localization and protrusion in this region, contributing
to migration against the direction of flow \citep{polacheck2014mechanotransduction}. 
This mechanism, which is dominant in 3D cell cultures, is similar to the mechanism underlying \textit{durotaxis}, where cells on a 2D substrate migrate in response to gradients in the mechanical stiffness of the substrate \citep{lo2000cell,duchez2019durotaxis}. Cell movement in response to gradients in cell-ECM adhesion-forces has been termed \textit{rheotaxis} in \cite{polacheck2014mechanotransduction}, but here we refer to it as \textit{tensotaxis} \citep{rosalem2020mechanobiological} in order to emphasize the role of fluid-induced stress (rather than velocity gradients) on this type of movement.
In addition to upstream directed movement induced by tensotaxis, autologous chemotaxis drives cell movement downstream. Here, the flow advects cell-secreted 
ligands, creating transcellular gradients of chemokines. The ligands bind to specific receptors on the cell surface, inducing cell polarization in the direction of higher chemokine concentrations and driving downstream, chemotactic migration. This autologous signaling mechanism has been observed by \cite{shields2007autologous}, where tumour cells have been shown to migrate downstream by binding self-secreted CCL21 ligands to the CCR7 receptors.

\begin{figure}[H]
\begin{centering}
\includegraphics[scale=1.2]{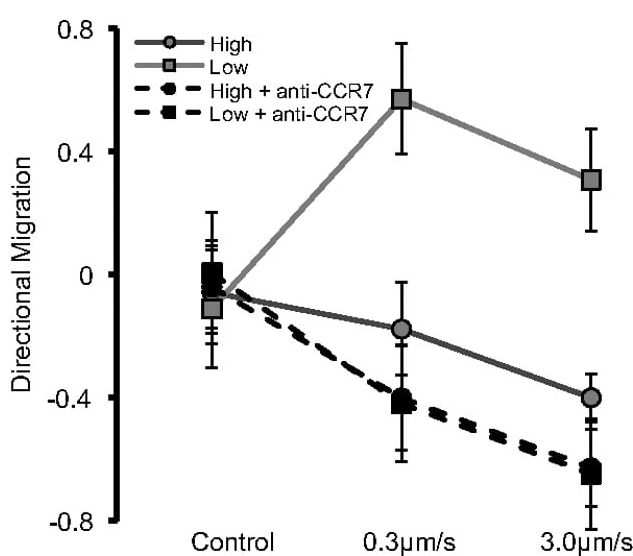}
\par\end{centering}
\caption{Experimental results, reproduced with permission from \cite{polacheck2011interstitial}, showing the directional migration score [positive (negative) -- most cells travel downstream (upstream), see \cite{polacheck2011interstitial} for details] as a function of the strength of interstitial flow and for different cell seeding densities (``high" and ``low" refer to seeding densities of $25 \times 10^4$ and $5 \times 10^4$ cells/mL, respectively). The dashed lines show that upstream migration prevails when CCR7 receptor signaling pathway is blocked, interrupting the downstream-oriented autologous chemotaxis.\label{fig:exp_results}}
\end{figure}

In experiments by \cite{polacheck2011interstitial}, cancer cells were seeded in a microfluidic channel and subject to fluid flow. The distribution of cell velocities was measured and the average migration direction (with respect to the flow direction) was evaluated; the local flow direction experienced by the cells was evaluated by numerically simulating the flow field in the microfluidic device. The results, reproduced in Fig.~\ref{fig:exp_results}, show that the dominant mode of migration switched between downstream (with the flow direction; positive values of Directional Migration in Fig.~\ref{fig:exp_results}) and upstream (against the flow direction; negative values of Directional Migration in Fig.~\ref{fig:exp_results}) as the cell density increased. 
However, when the CCR7 receptor signaling pathway was blocked, upstream migration was found to prevail regardless of the cell density, supporting the observations by \cite{shields2007autologous} regarding CCR7-dependent, downstream-oriented, autologous chemotaxis.
Additionally, for all of the experimental curves shown in Fig.~\ref{fig:exp_results}, an increase in the interstitial flow led to a higher tendency of the cells to migrate upstream. 
These results motivate the question of how different properties of cells, and the mechanochemical landscape they sense, affect their migration directions. In this paper, we show how mathematical modelling can shed light on the mechanisms regulating the direction of collective cell migration in a flow as system parameters vary. 

Models of chemotactic migration go back to the highly influential work of \cite{keller1971model}. More recently, chemotaxis has been considered in
the context of two-phase cellular tissue models \citep{byrne2004new,green2018pattern}, by formulating mass and momentum balance of the cell and fluid phases, coupled to the transport equation for chemoattractant propagation in the fluid phase. The ability of multiphase models to incorporate coupled interactions between cells, fluid, and chemoattractants makes them a natural framework for describing the mechanisms involved in mechanochemical transduction of cells subject to interstitial fluid flow.

While models for chemotaxis are prevalent [see the extensive review by \cite{arumugam2021keller}],
models for tensotaxis are less common. In a recent related work \citep{painter2021impact}, a generalised Keller-Segel model was applied to study the combined effect of rheotaxis (directed movement in response to flow velocity field) and chemotaxis on the aggregation of swimming organisms. However, in \cite{painter2021impact} a single-phase model was formulated, thus the coupled interactions of the organisms and fluid were not considered.
Evje and coworkers \citep{waldeland2018competing, evje2020mathematical} have formulated a multiphase
model which combined the competing mechanisms of downstream-oriented
autologous chemotaxis and upstream force, introduced \textit{ad hoc}
by inverting the direction of fluid drag force acting on the cells.
Their models have been successful in reproducing the transition between downstream and upstream migration as the cell's volume fraction increases, as
was observed in \cite{polacheck2011interstitial}. However, the heuristic assumption of inverting the direction of drag force does not explain the
mechanisms underlying this type of migration and does not allow any
generalisation of the model to more complicated scenarios where different
sources of mechanical stimulus exist. More recently, \cite{rosalem2020mechanobiological} derived a single-phase model for the tensotactic migration of cells, where the cell flux was assumed to be proportional to the transcellular pressure gradient. They verified that, in the presence of flow, this mechanism leads to upstream migration of cells; however, they did not consider the opposing effect of chemotactic migration. Consequently, it remains to be established what parameters (other than cell
volume fraction) affect the direction of cell migration and what
parameter regimes support downstream, rather than upstream, cell migration.

An additional drawback of existing macroscopic multiphase models for cell migration [e.g., \cite{byrne2004new, waldeland2018competing}] is that directed-migration is modelled as an internal force exerted by the cells on their self (source terms in the cell momentum balance). Therefore, the cell speed increases with the strength of the stimulus. This contradicts some experimental findings, which show that individual cells bias their directionality in response to the external cues, but their speed of migration is not correlated with their directionality \citep{polacheck2011interstitial, nam2020cancer,duchez2019durotaxis}.

The goal of the present work is to derive a two-phase model
for cell migration subject to flow-induced mechanochemical stimuli. The chemotactic and tensotactic cues are viewed as external
signals that bias the probability that a cell moves in a certain direction, while the magnitude of its speed remains constant. This situation resembles the kinetic model developed by \cite{hillen2006m5} to describe contact guidance of cell migration along the ECM fiber network. In that model, \cite{hillen2006m5} used a \emph{transition-probability function} (TPF) to describe the velocity-jump process of the cells, biased by the structure of the ECM network. The use of a TPF was originally introduced in pioneering work by \cite{othmer1988models} and has been used since to describe many aspects of directed cell migration [see the review by \cite{perthame2019kinetic} and references cited therein].
Building on these previous works, we propose a mesoscopic-kinetic equation to describe the evolution of the cells' probability density function in response to the external stimulus induced by fluid flow; here, the stimulus induces a cell-velocity-jump via a specifically tailored TPF. We then apply velocity-space averaging to derive continuum equations for the spatio-temporal
evolution of the cell volume fraction and flux, in response to forcing
terms depending on the local direction and magnitude of the mechanochemical stimulus. Using a combination of numerical simulations and asymptotic analysis, we delineate the parameter regimes for which cell migration transitions from downstream to upstream.

The remainder of the manuscript is structured as follows. In the Methods section
we introduce our two-phase model for cell
migration in response to flow-induced mechanochemical stimuli; we then introduce the one-dimensional model problem that is the focus of this paper, and use asymptotic methods to derive the critical conditions for transition between downstream
and upstream migration in the limit of small stimulus. In the Results and Discussion section we present numerical results describing the spatio-temporal dynamics of the cell layer for different parameter regimes supporting different modes of migration; then, we compare predictions from the asymptotic analysis with numerical results derived from the full model regarding the conditions under which migration switches between downstream and upstream regimes. In the concluding section we summarise our findings and outline possible directions for future research.

\section{Methods}

\subsection{Model formulation }

In this section we introduce a model for cell migration in the presence
of interstitial fluid flow, motivated by \textit{in vitro} experiments which show that interstitial flow can
induce mechanochemical stimuli biasing the direction of cell migration \citep{polacheck2011interstitial,shields2007autologous}.
In the present model we view the mechanochemical cues as external
signals that regulate the probability that the cells move in a certain direction, and assume that the magnitude of the cell speed remains
constant. In more detail, we seek to formulate a kinetic model for the probability density
function, $f(\mathbf{x},t,\boldsymbol{\xi})$, that the velocity of a cell
in a neighbourhood of spatial position $\mathbf{x}$,
at time $t$, has orientation vector $\boldsymbol{\xi}$. In the following subsection we formulate a dimensional model in an arbitrary number of dimensions;
in the subsequent subsection we consider a simplified one-dimensional version of the model and then non-dimensionalise the governing equations using the
characteristic scales of the system which we introduce therein.

\subsubsection{Probabilistic model for cell migration}

We consider a mesoscopic-kinetic model for the cells' probability density function, $f(\mathbf{x},t,\boldsymbol{\xi})$. We assume that the cells travel at a drift velocity $U_{c}\boldsymbol{\xi}$,
where the cell speed, $U_{c}$, is constant and $\boldsymbol{\xi}$
is a unit vector representing the cell-velocity orientation, which evolves in response to the external stimulus sensed by the cells. This change in cell-velocity orientation is modeled using a velocity-jump process induced by a transition probability function, $F(\mathbf{x},t,\boldsymbol{\xi})$.  We assume further that the cells perform an unbiased random walk (modeled as a microscopic space-jump process) superimposed
on their directed movement. Accordingly, we model the change in the
cell's probability density function, $f(\mathbf{x},t,\boldsymbol{\xi})$,
using the following mesoscopic-kinetic model:
\begin{equation}
\frac{\partial f}{\partial t} + U_c\boldsymbol{\xi}\cdot \boldsymbol{\nabla} f = \frac{1}{\tau}\left[ \int_{\boldsymbol{\xi}' \in \mathcal{V}}{F(\mathbf{x},t,\boldsymbol{\xi})f(\mathbf{x},t,\boldsymbol{\xi}')d\boldsymbol{\xi}'} -f\right] + D_c\nabla^2f,\label{eq:f_BGK}
\end{equation}
where $D_{c}$ denotes diffusivity due to the unbiased random motion; $\boldsymbol{\xi}$ and $\boldsymbol{\xi}'$ mark the current and previous cell-velocity orientation, respectively. The transition probability, $F$,
represents the rate at which the orientation vector changes, and biases
the probability density function in the direction of the stimulus; the constant, $\tau$,
represents the relaxation time over which the cell responds to the external signal. The integration in Eq.~(\ref{eq:f_BGK}) is carried out over all cells at $(\mathbf{x},t)$ [i.e., over all possible (previous) velocity orientations, $\boldsymbol{\xi}' \in \mathcal{V}$, where $\mathcal{V}$ is the set of vectors pointing from the origin to the surface of a unit sphere].
We assume that $F=F(\mathbf{x},t,\boldsymbol{\xi})$ is only a function of the current cell velocity (after jump) and not a function of the previous velocity (before jump). Therefore, Eq.~(\ref{eq:f_BGK}) can be written as
\begin{equation}
\frac{\partial f}{\partial t} + U_c\boldsymbol{\xi}\cdot \boldsymbol{\nabla} f =  \frac{1}{\tau}\left[\frac{\phi}{V_c}F-f\right] + D_c\nabla^2f,
\label{eq:f_BGK_2}
\end{equation}
where we assume that all cells have volume, $V_{c}$. Thus, we define the cell volume fraction as
\[
\phi = V_c \int_{\boldsymbol{\xi} \in \mathcal{V}}{f(\mathbf{x},t,\boldsymbol{\xi})d\boldsymbol{\xi}}.
\]

We note that the kinetic formulation in Eq.~(\ref{eq:f_BGK}) contains terms that arise from both space- and velocity-jump processes. Since this approach is non-standard, we now include more explanatory details. The velocity-jump process induced by the transition probability, $F$, will introduce a diffusion term at the second moment of $f$. However, the cell ``diffusivity'' will then scale as $U_c^2 \tau$ [see, for example, \cite{hillen2013transport}]. Under this scaling, we estimate that, for tumour cell migration, the diffusion term will be subdominant to the drift term. We note that this may contradict some experimental results on tumour cell migration for which the diffusive displacement is typically larger than the directed displacement [e.g., \cite{nam2020cancer,haessler2012migration}]. In contrast to, for example, a run-and-tumble process that characterises some bacteria, with cell migration there is no particular reason to assume {\em a priori} that the Brownian and directed motion of cells share similar speeds, particularly when the latter involves sensing which may reduce its speed relative to the speed of random motion. By assuming two separate stochastic processes [i.e., an isotropic space-jump process and an anisotropic (stimulus-induced) velocity-jump process], we can control the scales of both diffusive and directed motion and, thus, consider their effects separately. \cite{painter2013mathematical} used a transition probability that includes both isotropic and anisotropic parts to model the two separate processes; however, we prefer to use an isotropic space-jump process because it allows for simpler closure of the macroscopic model.

In Eq.~(\ref{eq:f_BGK}), cell-cell interactions and cell-volume exclusion are neglected. As such, the model is suitable
to describe situations in which the cell volume fraction takes low to moderate values. While the diffusion term in Eq.~(\ref{eq:f_BGK})
may mimic the effect of intercellular repulsion, other phenomena related
to collective cell migration \citep{tambe2011collective,angelini2011glass} are neglected in our model formulation, in order to focus attention on the way in which flow-induced
stimuli direct cell migration. In Eq.~(\ref{eq:f_BGK}) we also neglect cell proliferation and death since we aim to model migration dynamics at much shorter time scales.

We define a stimulus vector, $\mathbf{s}=|\mathbf{s}|\boldsymbol{\eta}$, where
$\boldsymbol{\eta}$ is a unit vector and $|\mathbf{s}|$ represents its magnitude, such that $F$ is maximized when the cell velocity and stimulus vector are aligned, and $F$ decreases monotonically as the
angle between $\boldsymbol{\xi}$ and $\boldsymbol{\eta}$ increases.
A simple functional form that captures this behaviour is given by
\begin{equation}
F\left(\mathbf{s}(\mathbf{x},t),\boldsymbol{\xi}\right)=A\exp\left[\mathbf{s}\cdot(\boldsymbol{\xi}-\boldsymbol{\eta})\right],\label{eq:F_first}
\end{equation}
where the normalisation factor $A$ is chosen so that $F$ satisfies
\begin{equation}
\int_{\boldsymbol{\xi} \in \mathcal{V}}{F d\boldsymbol{\xi}} = 1,
\label{eq:F_normalisation}
\end{equation}
to ensure conservation of mass.
Equation~(\ref{eq:F_first})
states that the velocity-jump process depends
solely on the local stimulus vector. 

We define $\theta\in[0,\pi]$ as the angle between $\boldsymbol{\xi}$
and $\boldsymbol{\eta}$, such that the three-dimensional integration $d \boldsymbol{\xi} = \sin\theta d\theta d \varphi$, where $\varphi \in [0,2\pi]$ is the polar angle in the plane perpendicular to $\boldsymbol{\eta}$. Therefore, the integral in Eq.~(\ref{eq:F_normalisation}) reads
\begin{equation}
\int_{\boldsymbol{\xi} \in \mathcal{V}}{ Fd\boldsymbol{\xi}}=2\pi A\int_{0}^{\pi}{\exp\left[|\mathbf{s}|(\cos\theta-1)\right]\sin\theta d\theta}=2\pi A \frac{(1-e^{-2|\mathbf{s}|})}{|\mathbf{s}|}.
\label{eq:F_d_xi}
\end{equation}
Combining Eqs.~(\ref{eq:F_normalisation}) and~(\ref{eq:F_d_xi}) we have
\begin{equation}
A = \frac{|\mathbf{s}|}{2\pi (1-e^{-2|\mathbf{s}|})}.
\end{equation}
Accordingly, $F$ is given by
\begin{equation}
F=\frac{|\mathbf{s}|}{2\pi (1-e^{-2|\mathbf{s}|})}\exp\left[\mathbf{s}\cdot(\boldsymbol{\xi}-\boldsymbol{\eta})\right].\label{eq:F_final}
\end{equation}

We multiply Eq.~(\ref{eq:f_BGK_2}) by
$V_{c}$ and integrate over all possible cell-velocity orientations, $\boldsymbol{\xi} \in \mathcal{V}$. Then, together with the normalisation condition of $F$ [Eq.~(\ref{eq:F_normalisation})], we have the macroscopic cell conservation equation
\begin{equation}
\frac{\partial\phi}{\partial t}+\boldsymbol{\nabla}\cdot\boldsymbol{\psi}=D_{c}\nabla^{2}\phi,
\label{eq:cell_conserv}
\end{equation}
where the cell flux, $\boldsymbol{\psi}$, is given by
\[
\boldsymbol{\psi}=V_{c}U_{c}\int_{\boldsymbol{\xi} \in \mathcal{V}}\boldsymbol{\xi}fd\boldsymbol{\xi}.
\]

In order to close the model we require an additional
equation for $\boldsymbol{\psi}$, and a natural choice is a momentum balance equation.
In order to derive the momentum balance, we multiply Eq.~(\ref{eq:f_BGK})
by $V_{c}U_{c}\boldsymbol{\xi}$ and integrate over all possible velocity orientations to obtain
\begin{equation}
\frac{\partial\boldsymbol{\psi}}{\partial t}+U_c^2 V_c \boldsymbol{\nabla}\cdot \int_{\boldsymbol{\xi} \in \mathcal{V}} \boldsymbol{\xi}\boldsymbol{\xi}^T f d\boldsymbol{\xi} =\frac{1}{\tau}\left(\phi U_{c}\int_{\boldsymbol{\xi} \in \mathcal{V}} F\boldsymbol{\xi}d\boldsymbol{\xi}-\boldsymbol{\psi}\right)+D_{c}\nabla^{2}\boldsymbol{\psi}.\label{eq:momentum_cell_first}
\end{equation}

The system of macroscopic equations~(\ref{eq:cell_conserv}) and~(\ref{eq:momentum_cell_first}) is not closed due to the presence of the second-order moment in the left-hand-side of Eq.~(\ref{eq:momentum_cell_first}). There is a wide body of literature concerning the formal derivation of macroscopic equations from kinetic models [see, for example, \cite{hillen2013transport} for closures of cell migration models and \cite{sone2007molecular} for derivation of hydrodynamic equations from gas kinetic theory]. However, for the present case of tumour cell migration, a simpler way to close the model is facilitated by the following dimensional analysis. We assume that the characteristic length scale of the system is $L$ and the characteristic time scale is the relaxation time, $\tau$. The characteristic scale of the cell flux is $\boldsymbol{\psi} \sim U_c$. Applying this scaling to Eq.~(\ref{eq:momentum_cell_first}) we have
\begin{equation}
\frac{\partial\boldsymbol{\widetilde{\psi}}}{\partial t}+\frac{U_c V_c \tau}{L} \boldsymbol{\widetilde{\nabla}}\cdot \int_{\boldsymbol{\xi} \in \mathcal{V}} \boldsymbol{\xi}\boldsymbol{\xi}^T f d\boldsymbol{\xi}=\phi\int_{\boldsymbol{\xi} \in \mathcal{V}} F\boldsymbol{\xi}d\boldsymbol{\xi}-\boldsymbol{\widetilde{\psi}}+\frac{D_{c} \tau}{L^2}\widetilde{\nabla}^{2}\boldsymbol{\widetilde{\psi}},\label{eq:momentum_cell_scaled}
\end{equation}
where tildes denote nondimensional variables and operators. From Eq.~(\ref{eq:f_BGK_2}), we estimate that the characteristic magnitude of $f$ is $F\phi/V_c$ (this is the equilibrium probability in the limit of small cell speed and diffusivity). Assigning $f \sim F\phi/V_c$ to the second-order moment in Eq.~(\ref{eq:momentum_cell_scaled}) we have
\begin{equation}
 V_c\int_{\boldsymbol{\xi} \in \mathcal{V}} \boldsymbol{\xi}\boldsymbol{\xi}^T f d\boldsymbol{\xi} \sim\frac{|\mathbf{s}|}{2\pi (1-e^{-2|\mathbf{s}|})}\int_{\boldsymbol{\xi} \in \mathcal{V}} \boldsymbol{\xi}\boldsymbol{\xi}^T\exp\left[\mathbf{s}\cdot(\boldsymbol{\xi}-\boldsymbol{\eta})\right] d\boldsymbol{\xi} \le O(1),
 \label{eq:second_moment}
\end{equation}
where it can readily be shown that the absolute magnitude of the components of the tensor in Eq.~(\ref{eq:second_moment}) are bounded such that they lie in the interval $[0,1]$ for all $\boldsymbol{s}$. Therefore, we assume that 
\begin{equation}
\frac{U_c V_c \tau}{L} \boldsymbol{\widetilde{\nabla}}\cdot \int_{\boldsymbol{\xi} \in \mathcal{V}} \boldsymbol{\xi}\boldsymbol{\xi}^T f d\boldsymbol{\xi} \sim O\left(\frac{U_c \tau}{L}\right).
\label{eq:O_2nd_moment}
\end{equation}
For tumour cell migration we estimate that the characteristic cell speed is $U_{c}\sim10\,\mathrm{\mu m/h}$
\citep{polacheck2011interstitial,nam2020cancer} and that the relaxation time is of the order of minutes to a few hours $\tau \lesssim 1\,\mathrm{h}$ \citep{wyatt2016question}. In order for the macroscopic model to be valid, we assume that the characteristic length scale is much larger than the dimensions of a single cell, $L \gg 10 \mathrm{\mu m/h}$. These characteristic scales lead to the conclusion that
\[
\frac{U_c \tau}{L} \ll 1
\]
and, therefore, that the second-order-moment term [Eq.~(\ref{eq:O_2nd_moment})] can generally be neglected in Eq.~(\ref{eq:momentum_cell_first}) in the context of tumour cell migration. We expect that a \emph{Chapman-Enskog} expansion in the small parameter $U_c \tau / L$, similar to that carried out by \cite{hillen2006m5}, would result in $O(U_c \tau / L)$ diffusive correction terms. However, such an expansion is beyond the scope of the current manuscript.

We note that in this model the length scale parameter, $U_c \tau$, represents the mean-free path of the cells: it describes the characteristic distance a cell covers before changing direction. Therefore, the assumption that the parameter $U_c \tau / L \ll 1$ means that in the macroscopic limit, cells undergo many velocity jumps along the macroscopic length scale. This parameter emerges as the equivalent to the Knudsen number in the kinetic theory of gases \citep{sone2007molecular}, which represents the ratio of the molecular mean-free path to the macroscopic length scale. Similarly to our model, in the kinetic theory of gases the derivation of macroscopic hydrodynamic equations from the kinetic Boltzmann equation is facilitated in the limit of small Knudsen number.

While it is possible to continue with the nondimensional formulation given by Eq.~(\ref{eq:momentum_cell_scaled}), we now return to the dimensional formulation and postpone non-dimensionalisation to the section where we introduce the model-problem setup.
Reverting to the dimensional formulation in Eq.~(\ref{eq:momentum_cell_first}) and evaluating the transition-probability-integral on the right-hand-side, we have
\begin{equation}
\int_{\boldsymbol{\xi} \in \mathcal{V}} F\boldsymbol{\xi}d\boldsymbol{\xi} = \frac{|\boldsymbol{s}|}{2\pi(1-e^{-2|\boldsymbol{s}|})}\int_{\boldsymbol{\xi} \in \mathcal{V}}\boldsymbol{\xi} \exp\left[\boldsymbol{s}\cdot(\boldsymbol{\xi}-\boldsymbol{\eta})\right]d\boldsymbol{\xi}  = \left(\mathrm{coth}(|\boldsymbol{s}|)-|\boldsymbol{s}|^{-1}\right)\boldsymbol{\eta},
\label{eq:integral_momentum}
\end{equation}
where the components in directions perpendicular to $\boldsymbol{\eta}$ vanish due to symmetry.

Finally, neglecting the second-order moment term in Eq.~(\ref{eq:momentum_cell_first}) and substituting from Eq.~(\ref{eq:integral_momentum}) we have
\begin{equation}
\frac{\partial\boldsymbol{\psi}}{\partial t}=\frac{1}{\tau}\left[\phi U_{c}\left(\mathrm{coth}(|\boldsymbol{s}|)-|\boldsymbol{s}|^{-1}\right)\boldsymbol{\eta}-\boldsymbol{\psi}\right]+D_{c}\nabla^{2}\boldsymbol{\psi}.
\label{eq:momentum_cell_second}
\end{equation}
We note that the cell flux source term in Eq.~(\ref{eq:momentum_cell_second}) is in the direction of the stimulus and is monotonically increasing with the stimulus magnitude, $|\boldsymbol{s}|$ (vanishes as $|\boldsymbol{s}| \rightarrow 0$ and attains a maximal value of $\phi U_c$ as $|\boldsymbol{s}| \rightarrow \infty$). This source term acts as an effective ``force'' driving cells in the direction of the stimulus.
In what follows, we propose
a constitutive model for the stimulus, $\mathbf{s}$, which depends on the local mechanochemical cues sensed by the cells.

\subsubsection{Constitutive model for the mechanochemical stimulus}

Cancer cells react to a variety of chemical and mechanical stimuli. We introduce a stimulus potential, $\Phi$, such that
\begin{equation}
\mathbf{s}=-l_{c}\boldsymbol{\nabla}\Phi,
\label{s_potential}
\end{equation}
where $l_{c}$ is a constant length which should be at the scale of
the cell length.
We consider two stimulus potentials:

\paragraph*{(I) Chemotaxis} Binding of ligands to specifc receptors on the membrane of cancer cells can polarise their movement in the direction of larger concentration of these ligands, leading to effective chemotactic migration \citep{roussos2011chemotaxis}. We model this process
by assuming that the potential of the chemotactic stimulus is proportional
to a chemokine concentration, $a$,

\begin{equation}
\Phi_{C}=-\chi a,
\end{equation}
where the constant $\chi$ represents the chemotactic potential
per unit concentration.

\paragraph*{(II) Tensotaxis} Cells respond to local stress by biasing their movement in the direction of larger tension in their cell-ECM connections
\citep{polacheck2011interstitial,polacheck2014mechanotransduction}. When
cells embedded in a three-dimensional matrix are subject to interstitial
flow, the cell response is usually stimulated by 
increased fluid pressure
at the upstream part (the part facing the flow) of the cell, which causes the cell to generate tensile ECM-adhesion forces in this region (and compressive forces in downstream region) resisting the flow-induced drag force. In turn,
the localized tension at the upstream part of the cell polarises its movement in the direction of larger pressure \citep{polacheck2014mechanotransduction}.
For simplicity, we model this process by viewing the tensotactic stimulus experienced by the cells as a potential
which is proportional to the stresses acting on the cell in the direction normal to the cells' outer surface:
\begin{equation}
\Phi_{T}=-\varpi\sigma^{\mathrm{ext}}_{nn}.\label{eq:tensotaxis}
\end{equation}
In Eq.~(\ref{eq:tensotaxis}), $\varpi$ represents the strength of the tensotactic potential
per unit stress, and $\sigma^{\mathrm{ext}}_{nn} \equiv \mathbf{n}^T \boldsymbol{\sigma}^{\mathrm{ext}} \mathbf{n}$ is the extracellular stress acting on the cell in the direction normal to its outer surface, where $\boldsymbol{\sigma}^{\mathrm{ext}}$ is the extracellular-stress tensor and $\mathbf{n}$ is a unit vector normal to the cell surface. We note here that, in more general cases, the cells may be subject to other external
stresses, such as shear stresses \citep{ostrowski2014microvascular} that stimulate tensotaxis.
Typically, when cells are embedded in a 3D matrix and subject to fluid flow, the dominant stress
they experience is due to fluid pressure \citep{polacheck2011interstitial}.
Therefore, in this work we assume that Eq.~(\ref{eq:tensotaxis})
can be simplified to read

\begin{equation}
\Phi_{T}=-\varpi p,
\end{equation}
where $p$ is the interstitial fluid pressure.

Finally, we write the total potential, $\Phi$, as the sum of the
chemotactic and tensotactic potentials, so that $\Phi = \Phi_{C}+\Phi_{T}$. Then, Eq.~(\ref{s_potential}) becomes
\begin{equation}
\mathbf{s}=l_{c}\varpi\boldsymbol{\nabla}p+l_{c}\chi\boldsymbol{\nabla}a.\label{eq:s_first}
\end{equation}

In order to close the model we introduce equations for $p$ and $a$. In what follows we formulate the governing equations of the flow dynamics in the two-phase cell-fluid mixture, such that the fluid pressure and the concentration of the flow-advected chemokine  can be evaluated.

\subsubsection{Interstitial flow dynamics}

We make a no-voids assumption for the cell-fluid mixture such that the volume fraction of the fluid phase is given by $1-\phi$. Then, the mass conservation equation of the fluid phase can be written as
\begin{equation}
\frac{\partial(1-\phi)}{\partial t}+\boldsymbol{\nabla}\cdot\left[(1-\phi)\mathbf{u_{f}}\right]=0,\label{eq:fluid_conserv}
\end{equation}
where $\mathbf{u_{f}}$ is the fluid velocity. Combining Eq.~(\ref{eq:fluid_conserv}) with Eq.~(\ref{eq:cell_conserv}) we have
\begin{equation}
\boldsymbol{\nabla}\cdot\left[\boldsymbol{\psi} + (1-\phi)\mathbf{u_{f}}\right]=D_c\nabla^2\phi.\label{eq:no_void}
\end{equation}
Assuming the fluid flux is much larger than the cell flux, as is usual for biological tissues,
\begin{equation}
\boldsymbol{\psi}-D_c\boldsymbol{\nabla}\phi \ll (1-\phi)\mathbf{u_{f}},
\end{equation}
we can simplify Eq.~(\ref{eq:no_void}) to
\begin{equation}
\boldsymbol{\nabla}\cdot\left[(1-\phi)\mathbf{u_{f}}\right]=0.\label{eq:no_void_simplified}
\end{equation}

We proceed by assuming that the momentum equation of the fluid phase
involves a balance between the drag force exerted by the cells
and the pressure gradient (for simplicity, we neglect intra-phase
viscous stresses). This balance can be written as a Darcy-type equation

\begin{equation}
\mathbf{u_{f}}-\mathbf{u^{\mathrm{avg}}_{c}}=-k_{H}g(\phi)\boldsymbol{\nabla}p,
\label{eq:Darcy}
\end{equation}
where $\mathbf{u^{\mathrm{avg}}_{c}}=\boldsymbol{\psi}/\phi$ is the average cell velocity, and $k_{H}$ represents the hydrodynamic conductivity (permeability divided
by viscosity).
In Eq.~(\ref{eq:Darcy}), $g(\phi)$ describes how the drag depends on the cell volume fraction, $\phi$. For simplicity, we use the popular Carman-Kozney relation \citep{swartz2007interstitial} so that
\begin{equation}
g(\phi)=\frac{(1-\phi)^{3}}{\phi^{2}}.\label{eq:Carman-Kozney}
\end{equation}
We chose the Carman-Kozney model since it is arguably the simplest isotropic model for hydraulic permeability that includes the effect of cell volume fraction. The dependence on the volume fraction is important as it ensures that the tensotactic stimulus increases as the cell volume fraction increases, as has been observed experimentally by \cite{polacheck2011interstitial}.

In Eq.~(\ref{eq:Darcy}) we can neglect the cell velocity with respect to the fluid velocity in to obtain
\begin{equation}
\mathbf{u_{f}}=-k_{H}g(\phi)\boldsymbol{\nabla}p.\label{eq:Darcy_2}
\end{equation}
Substituting from Eqs.~(\ref{eq:Carman-Kozney}) and~(\ref{eq:Darcy_2}) to Eq.~(\ref{eq:s_first}), we can
write the equation for the stimulus vector as
\begin{equation}
\mathbf{s}=-\frac{l_{c}\varpi}{k_{H}}\frac{\phi^{2}}{(1-\phi)^{3}}\mathbf{u_{f}}+l_{c}\chi\boldsymbol{\nabla}a.\label{eq:s_second}
\end{equation}

Finally, we model the
evolution of the chemokine concentration, $a$, using a reaction-advection-diffusion equation. We assume that the chemokine
is secreted by the cells at a constant rate, $\beta_{p}$, and that $\beta_{d}$
is the rate (per unit concentration) at which it binds to receptors on the surface of the cells.
Under these assumptions we obtain the following equation for the chemokine concentration
\begin{equation}
\frac{\partial a}{\partial t}+\boldsymbol{\nabla}\cdot\left[a(1-\phi)\mathbf{u_{f}}\right]=\frac{\beta_{p}}{V_{c}}\phi-\frac{\beta_{d}}{V_{c}}\phi a+D\nabla^{2}a.\label{eq:c_dimesnional-1}
\end{equation}
where $D$ is the diffusion coefficient of the chemokine in the interstitial fluid.

Taken together, Eqs.~(\ref{eq:cell_conserv}),~(\ref{eq:momentum_cell_second}),~(\ref{eq:no_void_simplified}),~(\ref{eq:s_second}), and~(\ref{eq:c_dimesnional-1}) form a closed system for the cell volume fraction, $\phi$ and flux, $\boldsymbol{\psi}$, the cell stimulus, $\mathbf{s}$, fluid velocity, $\mathbf{u_{f}}$, and chemokine concentration, $a$.

\subsection{Model problem: cell layer subject to one-dimensional
flow\label{subsec:Model_problem_1d}}

\subsubsection{Formulation of a nondimensional 1D model}

\begin{figure}[H]
\begin{centering}
\includegraphics[scale=0.5]{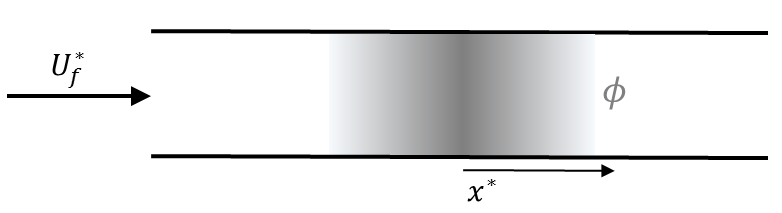}
\par\end{centering}
\caption{Schematic illustration of the one-dimensional model problem. \label{fig:schematic}}
\end{figure}

In this subsection we reduce the model developed in the previous subsection
to a one-dimensional model that describes the migration of
a population of cells (initially localized around a particular spatial position) in a long microfluidic channel (i.e., we neglect cell fluxes into, or out of, the channel edges).
This model will be used to provide a simple explanation of how and why the migration patterns of tumour cells change in response to changes in flow velocity and cell volume fraction observed by \cite{polacheck2011interstitial}. While the geometry of the microfluidic device in \cite{polacheck2011interstitial} is not exactly identical to a long channel, the fluid velocity in the cell-region of the experiments was primarily oriented in a single direction, along the axis of the channel. We, thus, view our 1D model as a reasonable approximation of the experimental setup.
We consider
a long channel which is aligned with the $x$-axis, in which an initial cell
layer is distributed normally around $x=0$ such that

\begin{equation}
\phi(x^{*},t^{*}=0)=\overline{\phi}\exp\left[-\left(\frac{x^{*}}{L^{*}}\right)^{2}\right].\label{eq:initial_phi}
\end{equation}
In Eq.~(\ref{eq:initial_phi}) and henceforth, we use asterisks to denote dimensional parameters,
and the constants $\overline{\phi}$ and $L^{*}$ represent typical values of the cells' initial volume fraction and
layer size, respectively. The cells are subject to fluid flow, where in the far-field as $|x^{*}|\gg L^{*}$ (i.e.,
in regions sufficiently far from the cell layer), the fluid velocity magnitude is $U_{f}^{*}$. A schematic of the one-dimensional model problem is illustrated in Fig.~\ref{fig:schematic}.

We now simplify
the equations derived in the Model Formulation subsection
to 1D Cartesian geometry form and non-dimensionalise them using the following
scaling: we normalize length by the characteristic length of the initial
distribution of cells, $L^{*}$; we scale the velocity by the far-field
fluid velocity, $U_{f}^{*}$; accordingly, time is normalized by $L^{*}/U_{f}^{*}$.
The chemokine concentration is scaled by its maximal equilibrium concentration,
$a_{\mathrm{eq}}^{*}=\beta_{p}^{*}/\beta_{d}^{*}$. The full set of independent and dependent nondimensional variables are given by
\begin{equation}
x = \frac{x^*}{L^*},\ t = \frac{t^*U_f^*}{L^*},\ \phi,\ \psi = \frac{\psi^*}{U_f^*}, \ u_f=\frac{u_f^*}{U_f^*}, \ \text{and}  \ a = \frac{a^*}{a_{\mathrm{eq}}^{*}}.
\end{equation}
Then, Eq.~(\ref{eq:cell_conserv}) in a 1D nondimensional form reads
\begin{equation}
\frac{\partial\phi}{\partial t}+\frac{\partial\psi}{\partial x}=\frac{1}{\mathrm{Pe}_{c}}\frac{\partial^{2}\phi}{\partial x^{2}},\label{eq:phi_1d_nondimensional}
\end{equation}
where $\psi$ is the $x$-component of $\boldsymbol{\psi}$ and
\[
\mathrm{Pe}_{c}=\frac{U_{f}^{*}L^{*}}{D_{c}^{*}}.
\]
Using physiologically relevant parameters we have $U_{f}^{*}\sim1\,\mathrm{\mu m/s}$
\citep{polacheck2011interstitial}, $L^{*}\sim 100\,\mathrm{\mu m}$, and $D^*_{c}\sim1000\,\mathrm{\mu m^{2}/h}$ \citep{marel2014flow}, such that
the interstitial fluid velocity is much larger than the diffusive
velocity of cells, i.e., $\mathrm{Pe}_{c}\gg1$. We expect, however, that
the cell flux will also be small, $\psi\ll1$, such that we cannot
neglect diffusive effects.

In one dimension it is easier to assume that the direction of the
stimulus is constant, $\boldsymbol{\eta}=\mathbf{\hat{x}}$, where $\mathbf{\hat{x}}$ is the unit vector in the positive $x$-direction.
Then, the stimulus vector is given by
\[
\boldsymbol{s}=s\mathbf{\hat{x}},
\]
where $s\in(-\infty,\infty)$ is the $x$-component of the stimulus vector.
We note that the stimulus vector is aligned with the $x$-axis because, within the 1D model, the only gradients of the macroscopic fields are in the $x$-direction. However, the microscopic velocity, $\boldsymbol{\xi}$, is still a three-dimensional vector, in accordance with its definition in Eq.~(\ref{eq:f_BGK}) \emph{et seq}.

Reducing Eq.~(\ref{eq:momentum_cell_second}) to 1D and non-dimensionalising we have
\begin{equation}
\frac{\partial\psi}{\partial t}=\frac{\mathcal{U}}{\mathcal{T}}\left[\mathrm{coth}(s)-s^{-1}\right]\phi-\frac{\psi}{\mathcal{T}}+\frac{1}{\mathrm{Pe}_{c}}\frac{\partial^{2}\psi}{\partial x^{2}},\label{eq:psi_1d_nondimensional}
\end{equation}
where
\[
\mathcal{T}=\frac{\tau^{*}U_{f}^{*}}{L^{*}}\;\;\text{and}\;\;\mathcal{U}=\frac{U_{c}^{*}}{U_{f}^{*}}.
\]
In Eq.~(\ref{eq:psi_1d_nondimensional}) we made use of the antisymmetry of the source term in Eq.~(\ref{eq:momentum_cell_second}):
\begin{equation}
\mathrm{coth}(s)-s^{-1} = \begin{cases}
-\left(\mathrm{coth}(|\mathbf{s}|)-|\mathbf{s}|^{-1} \right), & s \leq 0\\
\quad \ \mathrm{coth}(|\mathbf{s}|)-|\mathbf{s}|^{-1} \;\, , & s \geq 0.
\end{cases}  
\end{equation}

Since the cell velocity is much smaller than the fluid velocity we
will assume $\mathcal{U}\ll1$. The characteristic response time of
cells, $\tau^{*}$, is in the range of minutes to hours \citep{wyatt2016question} such that, based on the characteristic scales of length and fluid
velocity introduced above, we can estimate that $\mathcal{T}\sim 10-100$.

In order to evaluate the stimulus, $s$, we must solve for the
fluid velocity and chemokine gradient. Starting from the fluid velocity,
we consider the nondimensional 1D form of the cell-fluid mixture mass conservation Eq.~(\ref{eq:no_void_simplified})
\begin{equation}
\frac{\partial}{\partial x}\left[(1-\phi)u_{f}\right]=0.\label{eq:incom}
\end{equation}
Integrating Eq.~(\ref{eq:incom}) with respect to $x$ we have
\begin{equation}
u_{f}=\frac{1}{1-\phi}.\label{eq:u_f}
\end{equation}
where we have assumed that $u_{f}|_{x\rightarrow-\infty}=1$, in accordance with the dimensional boundary condition, $u^*_{f}|_{x^*\rightarrow-\infty}=U_f^*$.

The chemokine transport equation in a 1D nondimensional form then reads
\begin{equation}
\frac{\partial a}{\partial t}+\frac{\partial a}{\partial x}=\mathrm{Da}\,\phi\left(1-a\right)+\frac{1}{\mathrm{Pe}}\frac{\partial^{2}a}{\partial x^{2}},\label{eq:c_1d_nondimensional}
\end{equation}
where
\begin{equation}
\mathrm{Da}=\frac{L^{*}\beta_{d}^{*}}{U_{f}^{*}V_{c}^{*}}\;\;\text{and}\;\;\mathrm{Pe}=\frac{U_{f}^{*}L^{*}}{D^{*}}\label{eq:nondimensional_param_c}
\end{equation}
are the chemoattractant- Damkohler and Peclet numbers, respectively.
Using the physiologically relevant parameters of $U_{f}^{*}$ and
$L^{*}$ introduced above, together with characteristic diffusivity of chemokines $D^{*}\sim100\,\mathrm{\mu m^{2}/s}$
\citep{fleury2006autologous,bonneuil2022generation}, we can estimate that $\mathrm{Pe}\sim1$,
meaning that diffusive effects are likely to be important. In
the context of the present one-dimensional model, the diffusive terms act to smooth the chemokine gradient and, thereby, to reduce the magnitude of the
chemotactic cue in the centre of the cell layer (with subdominant
contributions at the edges of the domain). However, since $\mathrm{Pe} \ll \mathrm{Pe}_c$, we would need a very large domain in order to simulate, on the one hand, sufficiently large times to allow for cell migration while, on the other hand, avoiding boundary interactions of the chemokine at the channel edges.
Therefore, to simplify the numerical calculations, we choose to neglect the diffusive term in Eq.~(\ref{eq:c_1d_nondimensional}) and view the results as the purely-advective limit, bearing in mind that including diffusion would result in somewhat weaker chemotactic migration. 

Additionally, the time-derivative
of the chemokine is associated with changes in the cell volume fraction such that
\begin{equation}
\frac{\partial a}{\partial t}\sim\mathrm{Da}\frac{\partial\phi}{\partial t}\sim O\left(\mathrm{Da}\,\mathcal{U},\frac{\mathrm{Da}}{\mathrm{Pe}_{c}}\right)\ll1.
\end{equation}
Therefore, we can assume that the chemokine distribution is quasi-steady,
i.e., changes in the cell volume fraction lead to instantaneous adaption
of the chemokine distribution. Under these assumptions and together with the vanishing of the chemokine at the channel inlet, $a|_{x\rightarrow - \infty}=0$, we can simplify
Eq.~(\ref{eq:c_1d_nondimensional}) to
\begin{equation}
a(x,t)=1-\exp\left(-\mathrm{Da}\int_{-\infty}^{x}\phi(z,t)dz\right).\label{eq:c_1d_nondimensional_no_Pe}
\end{equation}

Substituting from Eqs.~(\ref{eq:u_f}) and~(\ref{eq:c_1d_nondimensional_no_Pe})
in the 1D form of Eq.~(\ref{eq:s_second}) we have
\begin{equation}
s=-\mathcal{\mathcal{K}}\frac{\phi^{2}}{(1-\phi)^{4}}+\mathcal{M}\mathrm{Da}\,\phi\exp\left(-\mathrm{Da}\int_{-\infty}^{x}\phi(z,t)dz\right),\label{eq:s_1d_final}
\end{equation}
where the nondimensional parameters
\[
\mathcal{\mathcal{K}}=\frac{l_{c}^{*}\varpi^{*}U_{f}^{*}}{k_{H}^{*}}\;\;\text{and}\;\;\mathcal{M}=\frac{l_{c}^{*}\chi^{*}a_{\mathrm{eq}}^{*}}{L^{*}},
\]
represent characteristic magnitudes of the tensotactic and
chemotactic stimuli, respectively. Then, the one-dimensional spatio-temporal
evolution of the cell volume fraction, $\phi$, and flux, $\psi$,
in response to interstitial fluid flow, can be solved using Eqs.~(\ref{eq:phi_1d_nondimensional})
and~(\ref{eq:psi_1d_nondimensional}), together with the constitutive
model for the cell stimulus in Eq.~(\ref{eq:s_1d_final}). We impose no flux boundary conditions at the far field, i.e.,
\begin{equation}
\frac{\partial \phi}{\partial x}=0 \ \ \text{and} \ \ \psi=0 \ \ \text{as} \ \ x \rightarrow \pm \infty.
\label{eq:BCs}
\end{equation}
In order to solve the system of equations given by Eqs.~(\ref{eq:phi_1d_nondimensional}),~(\ref{eq:psi_1d_nondimensional}), and~(\ref{eq:s_1d_final}), we use a semi-implicit
finite difference scheme; the $x$-derivatives are discretised using
a second-order central difference method on a uniform grid spanning the interval $[-X,X],$where $X \gg 1$ is sufficiently large that the far-field boundary conditions given in Eq.~(\ref{eq:BCs}) have negligible effect on the results. Advancing the system in time is achieved
using Euler's forward method. The above scheme is implemented in MATLAB. The code is available at the following GitHub repository: github.com/yaronbenami/cell\_migration.

\subsubsection{Downstream and upstream migrating populations}

An important goal of the present model is to 
identify parameter
regimes in which transitions between upstream and downstream migration
occur. While the sign of $\psi$ provides an indication
of the average direction of cell migration, changes in
the proportion of cells traveling upstream and downstream is more accurately
given by
\begin{equation}
\phi^{\mathrm{diff}}=V_c \left( \int_{\xi_{x}>0}fd\boldsymbol{\xi}-\int_{\xi_{x}<0}fd\boldsymbol{\xi} \right).\label{eq:phi_diff_definition}
\end{equation}
It is important to note here that, because we used a probabilistic approach to develop our model, we can 
derive $\phi^{\mathrm{diff}}$ from Eq.~(\ref{eq:phi_diff_definition}). This would not have been possible using a conventional multiphase model in which the average macroscopic variables are not explicitly related to microscopic velocity distributions.

Integrating Eq.~(\ref{eq:f_BGK_2}) with respect to $\boldsymbol{\xi}$
for $\xi_{x}>0$ and subtracting the integral of Eq.~(\ref{eq:f_BGK_2})
for $\xi_{x}<0$ we have
\begin{equation}
\frac{\partial\phi^{\mathrm{diff}}}{\partial t}+\frac{\partial\psi^{\mathrm{diff}}}{\partial x}=\frac{1}{\mathcal{T}}\left(\phi \mathrm{tanh}\left(\frac{s}{2}\right)-\phi^{\mathrm{diff}}\right)+\frac{1}{\mathrm{Pe}_{c}}\frac{\partial^{2}\phi^{\mathrm{diff}}}{\partial x^{2}},\label{eq:phi_diff}
\end{equation}
where
\begin{equation}
\psi^{\mathrm{diff}}=U_c V_c \left(\int_{\xi_{x}>0}\xi_{x}fd\boldsymbol{\xi}-\int_{\xi_{x}<0}\xi_{x}fd\boldsymbol{\xi} \right),
\end{equation}
and we note the source (sink) term due to cells changing their migration
direction from upstream to downstream (and vice versa). In order to
obtain an equation for $\psi^{\mathrm{diff}}$, we multiply Eq.~(\ref{eq:f_BGK_2})
by $U_c V_c \xi_{x}$, integrate with respect to $\boldsymbol{\xi}$ for $\xi_{x}>0$
and subtract the integral for $\xi_{x}<0$ to obtain
\begin{equation}
\frac{\partial\psi^{\mathrm{diff}}}{\partial t}=\frac{\mathcal{U}}{\mathcal{T}}\left(1-\frac{\mathrm{tanh}(s/2)}{s}\right)\phi-\frac{\psi^{\mathrm{diff}}}{\mathcal{T}}+\frac{1}{\mathrm{Pe}_{c}}\frac{\partial^{2}\psi^{\mathrm{diff}}}{\partial x^{2}}.\label{eq:psi_diff}
\end{equation}
With $\phi$ and $s$ determined by Eqs.~(\ref{eq:phi_1d_nondimensional}),~(\ref{eq:psi_1d_nondimensional}) and~(\ref{eq:s_1d_final}), we
can solve Eqs.~(\ref{eq:phi_diff}) and~(\ref{eq:psi_diff}) to determine $\phi^{\mathrm{diff}}$
and $\psi^{\mathrm{diff}}$. 

We define the total difference between downstream- and upstream-migrating
cells,
\begin{equation}
N^{\mathrm{diff}}(t)=\int_{-\infty}^{\infty}\phi^{\mathrm{diff}}(x,t)dx,\label{eq:int_phi_diff}
\end{equation}
and use this quantity as a metric to determine whether, at time $t$, there is a dominant
tendency for the cells to migrate downstream ($N^{\mathrm{diff}}(t)>0$) or upstream ($N^{\mathrm{diff}}(t)<0$). We note that by integrating Eq.~(\ref{eq:phi_diff}) with respect to $x$ we can derive an ODE for $N^{\mathrm{diff}}$
\begin{equation}
\frac{d N^{\mathrm{diff}}}{d t}=\frac{R(t)-N^{\mathrm{diff}}}{\mathcal{T}},\label{eq:N_diff_ODE}
\end{equation}
where we have assumed no cell flux as $x \rightarrow \pm \infty$ and  $R(t)$ is given by
\[
R(t) = \int_{-\infty}^{\infty}\phi  \mathrm{tanh}\left(\frac{s}{2}\right) dx.
\]

Finally, we define $\overline{\phi}_{\mathrm{cr}}(t)$ as the critical
value of $\overline{\phi}$ [the initial volume fraction at $x=0$, see Eq.~(\ref{eq:initial_phi})]
for which $N^{\mathrm{diff}}=0$ at time $t$, such that a transition in the
overall migration tendency occurs at time $t$ when $\phi(0,0)=\overline{\phi}_{\mathrm{cr}}$.
The experimental significance of $\overline{\phi}_{\mathrm{cr}}$ can be summarised as follows. Suppose we want to measure the dominant mode of cell migration after time $T$ has elapsed since the seeding of the cells. If the cells are seeded with an initial density $\overline{\phi}_{\mathrm{cr}}(T)$, then
equal proportions of cells will be migrating up- and down-stream at time $T$. If the cells are seeded at densities $\overline{\phi} > \overline{\phi}_{\mathrm{cr}}(T)$ then the dominant direction of cell migration at time $T$ will be upstream (and vice versa). Therefore, $\overline{\phi}_{\mathrm{cr}}$ is a measure that can be used to predict the expected dominant mode of migration at different times during the experiment.

\subsubsection{Asymptotic analysis of $\overline{\phi}_{\mathrm{cr}}$ in the limit of small
stimulus}

The strengths of the tensotactic and chemotactic stimuli are governed
by the nondimensional parameters $\overline{\phi}$ and $\mathrm{Da}$.
While the parameters $\mathcal{K}$ and $\mathcal{M}$ also affect
the value of $s$, we will show below that the transition between
downstream and upstream migration is dominated by the
parameter combinations that yield $s|_{x=0}=0$, such that only the
ratio, $\mathcal{K}/\mathcal{M}$, affects the value
of $\overline{\phi}_{\mathrm{cr}}$. 

In this section we estimate $\overline{\phi}_{\mathrm{cr}}$ in the limiting case for which $\overline{\phi},\mathrm{Da}\ll1$. 
In this limit, the order-of-magnitude of the stimulus in Eq.~(\ref{eq:s_1d_final})
scales as
\[
s\sim O\left(\overline{\phi}^{2},\mathrm{Da}\overline{\phi}\right)\ll1.
\]
In what follows we evaluate the asymptotic value of $s$ in the limit $s \ll 1$.
In Eq.~(\ref{eq:psi_1d_nondimensional}), in this limit, 
the cell advective flux, $\psi$, is much smaller than the diffusive flux, $\psi\sim\mathcal{U}\overline{\phi}s\ll\mathrm{Pe}_{c}^{-1}\overline{\phi}$.
Therefore, we may assume that, at leading order, the dynamics of the cell volume fraction
are governed by the  unsteady diffusion equation
\begin{equation}
\frac{\partial\phi}{\partial t}\approx\frac{1}{\mathrm{Pe}_{c}}\frac{\partial^{2}\phi}{\partial x^{2}}.\label{eq:cell_limit_diffusion}
\end{equation}
The solution of Eq.~(\ref{eq:cell_limit_diffusion}), together with
the initial condition
\[
\phi(x,0)=\overline{\phi}\exp\left(-x^{2}\right),
\]
and the far-field decay of $\phi$, is given by
\begin{equation}
\phi(x,t)\approx\frac{\overline{\phi}}{\sqrt{1+4t/\mathrm{Pe}_{c}}}\exp\left(-\frac{x^{2}}{1+4t/\mathrm{Pe}_{c}}\right).\label{eq:phi_limit}
\end{equation}

With $s\ll1$, at leading order, Eqs.~(\ref{eq:phi_diff}) and~(\ref{eq:psi_diff}) yield the following expressions for $\phi^{\mathrm{diff}}$ and $\psi^{\mathrm{diff}}$,
\begin{equation}
\phi^{\mathrm{diff}}=-\mathcal{T}\frac{\partial\psi^{\mathrm{diff}}}{\partial x}+O(s)\label{eq:phi_diff_limit}
\end{equation}
and
\begin{equation}
\psi^{\mathrm{diff}}=\frac{1}{2}\mathcal{U}\phi+O(s).\label{eq:psi_diff_limit}
\end{equation}
Combining Eqs.~(\ref{eq:phi_diff_limit}) and~(\ref{eq:psi_diff_limit}) we have
\begin{equation}
\phi^{\mathrm{diff}}=-\frac{1}{2}\mathcal{T}\mathcal{U}\frac{\partial\phi}{\partial x}+O(s).\label{eq:phi_diff_limit_2}
\end{equation}
Since in this limit $\phi^{\mathrm{diff}}$ and $\psi^{\mathrm{diff}}$ are, respectively, proportional to $\phi$ and $\partial \phi/\partial x$ , the unsteady diffusion operator, $\mathcal{L} \equiv \partial/\partial t - \mathrm{Pe}^{-1}\partial^2/\partial x^2$ was eliminated from Eqs.~(\ref{eq:phi_diff}) and~(\ref{eq:psi_diff}) by substituting from Eq.~(\ref{eq:cell_limit_diffusion}).

It can be readily verified from Eq.~(\ref{eq:phi_diff_limit_2}) that $\phi^{\mathrm{diff}}$ is antisymmetric with respect to $x=0$ [since $\phi$ maintains its
symmetry at this limit, see Eq.~(\ref{eq:phi_limit})]. Thus, the leading order term of $\phi^{\mathrm{diff}}$ does not contribute to the integral in Eq.~(\ref{eq:int_phi_diff}). We conclude that the contribution of the tensotactic and chemotactic stimuli to the
integral arises from the $O(s)$ terms
localized around $x=0$, at which $\partial\phi/\partial x$ vanishes.
Consequently, we can assume that the tendency towards upstream or
downstream migration is dominated by the stimulus value at
$x=0$. Assigning Eq.~(\ref{eq:phi_limit}) to Eq.~(\ref{eq:s_1d_final})
we have
\begin{equation}
s|_{x=0}\approx-\frac{\mathcal{K}}{(1+4t/\mathrm{Pe}_{c})}\frac{\overline{\phi}^{2}}{(1-\overline{\phi}/\sqrt{1+4t/\mathrm{Pe}_{c}})^{4}}+\frac{\mathcal{M}\mathrm{Da}\overline{\phi}}{\sqrt{1+4t/\mathrm{Pe}_{c}}}\exp\left(-\frac{\sqrt{\pi}\mathrm{Da}\overline{\phi}}{2}\right).\label{eq:s_limit}
\end{equation}
In Eq.~(\ref{eq:s_limit}), we retain terms that are subdominant
as $\mathrm{Da},\overline{\phi}\ll1$. While the higher order terms are not asymptotically valid (since we did not formally derive the next order correction terms), retaining them in our analysis was useful in order to capture the qualitative behaviour of $\overline{\phi}_{\mathrm{cr}}$
for non-small $\mathrm{Da}$ and $\overline{\phi}$ (see, for
example, the local maximum in $\overline{\phi}_{\mathrm{cr}}$ in Fig.~\ref{fig:critical_phi}
which is captured by the asymptotic expression). 

In the limit
when $\overline{\phi}\ll1$, $\mathcal{R}_{T/C}$, the ratio of the magnitudes of the tensotactic and chemotactic terms in Eq.~(\ref{eq:s_limit}), is proportional to $\overline{\phi}\ll1$:
\[
\mathcal{R}_{T/C} = \frac{\mathcal{K}\overline{\phi}}{\mathcal{M\mathrm{Da}}\sqrt{1+4t/\mathrm{Pe}_{c}}}.
\]
We conclude that, for sufficiently small cell volume fractions, there will always be a parameter combination such that $\mathcal{R}_{T/C}<1$, i.e., a dominant tendency towards downstream migration.
This result is consistent with the experimental findings of \cite{polacheck2011interstitial}
who observed that downstream migration becomes more dominant as the cell volume fraction decreases (see Fig.~\ref{fig:exp_results}). 

Equating Eq.~(\ref{eq:s_limit}) to
zero yields the following transcendental equation for $\overline{\phi}_{\mathrm{cr}}$:
\begin{equation}
\frac{\mathcal{K}}{\mathcal{M}\mathrm{Da}\sqrt{1+4t/\mathrm{Pe}_{c}}}\frac{\overline{\phi}_{\mathrm{cr}}\exp\left(\frac{\sqrt{\pi}\mathrm{Da}\overline{\phi}_{\mathrm{cr}}}{2}\right)}{(1-\overline{\phi}_{\mathrm{cr}}/\sqrt{1+4t/\mathrm{Pe}_{c}})^{4}}=1,\label{eq:phi_c_nonlinear}
\end{equation}
which depends on the nondimensional parameter groupings, $\mathcal{K}/\mathcal{M}$,
$\mathrm{Da}$ and $t/\mathrm{Pe}_{c}$, representing the relative strength of the tensotactic to chemotactic stimulus, the ratio of chemokine reaction and advection rates, and cell-diffusive time, respectively. The dependence of $\overline{\phi}_{\mathrm{cr}}$ on $t$ means that the critical value of the initial volume fraction for the transition from downstream to upstream migration depends on the time that has elapsed since the
cells were seeded.
This is because cell diffusion
reduces the value of $\phi|_{x=0}$ as $t$ increases.
Consequently, a transition to downstream migration
at sufficiently large $t$ will always occur, regardless of the value of the initial
volume fraction $\overline{\phi}$. Alternatively, by replacing $\overline{\phi}_{\mathrm{cr}} \rightarrow \overline{\phi}$ and $t \rightarrow t_{\mathrm{cr}}$ in Eq.~(\ref{eq:phi_c_nonlinear}), the equation could be interchanged to describe the critical time in which the transition takes place, $t_{\mathrm{cr}}$, as a function of the initial volume fraction $\overline{\phi}$.

\section{Results and discussion}

Table \ref{tab:Nondimensional-parameter-values} summarises the nondimensional parameter values used to generate model simulations. We will study the behaviour of cells for a range of values of the initial volume fraction, $\overline{\phi}$, and the Damkohler number, $\mathrm{Da}$. These parameters, together with the values of the tensotactic and chemotactic signal strengths, $\mathcal{K}$, and $\mathcal{M}$, respectively, govern the magnitude and direction of the stimuli. In this work we keep $\mathcal{K}$ and $\mathcal{M}$ constant and equal, and vary the values of $\overline{\phi}$ and $\mathrm{Da}$.
The cells' nondimensional velocity, $\mathcal{U}$ and their Peclet number, $\mathrm{Pe}_c$, are chosen to have physiologically relevant values and are fixed at these default values throughout the paper.
We will examine two physiologically relevant values of the cells' relaxation time, $\mathcal{T}$, corresponding to dimensional times of minutes and hours.

\begin{table}[H]
\begin{centering}
\begin{tabular}{|c|c|}

\hline
Parameter & Value\tabularnewline
\hline 
\rule{0pt}{3ex} 
$\overline{\phi}$ & 0.01-0.5\tabularnewline

$\mathrm{Da}$ & 0.01-10\tabularnewline

$\mathcal{U}$ & 0.003 (a)\tabularnewline

$\mathrm{Pe}_{c}$ & 300 (b)\tabularnewline

$\mathcal{T}$ & 10, 100 (c)\tabularnewline
 
$\mathcal{K}$ & 10\tabularnewline

$\mathcal{M}$ & 10\tabularnewline
\hline
 
\end{tabular}
\par\end{centering}
\medskip{}

\begin{adjustwidth}{100pt}{100pt}

\begin{enumerate}[label=(\alph*)]
\item \footnotesize{Assuming dimensional fluid velocity, $U_f^* \sim 1 \ \mathrm{\mu m/s}$, this gives a dimensional cell speed, $U_c^* \sim 10 \ \mathrm{\mu m/h}$ \citep{polacheck2011interstitial}}
\item \footnotesize{Assuming  $U_f^* \sim 1 \ \mathrm{\mu m/s}$ and $L^* \sim 100 \ \mathrm{\mu m}$, this gives a dimensional cell diffusivity, $D_c^* \sim 1000 \ \mathrm{\mu m^2/h}$ \citep{marel2014flow}}
\item \footnotesize{See discussion after Eq.~(\ref{eq:psi_1d_nondimensional}) \textit{et seq.}}
\end{enumerate}

\end{adjustwidth}

\caption{Nondimensional parameter values\label{tab:Nondimensional-parameter-values}}

\end{table}

\begin{figure}[hbt!]
\includegraphics[scale=0.31]{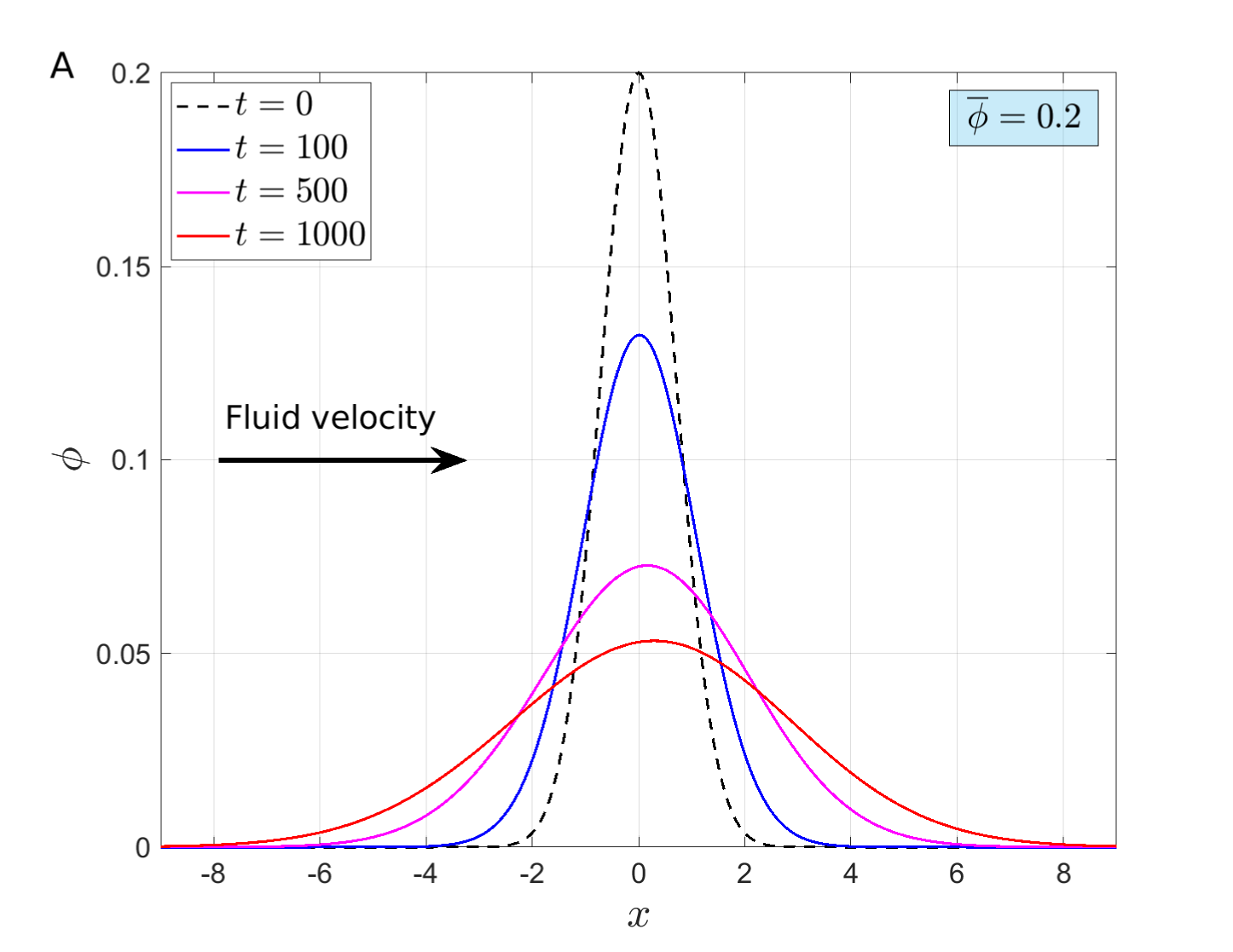}\hfill{}\includegraphics[scale=0.31]{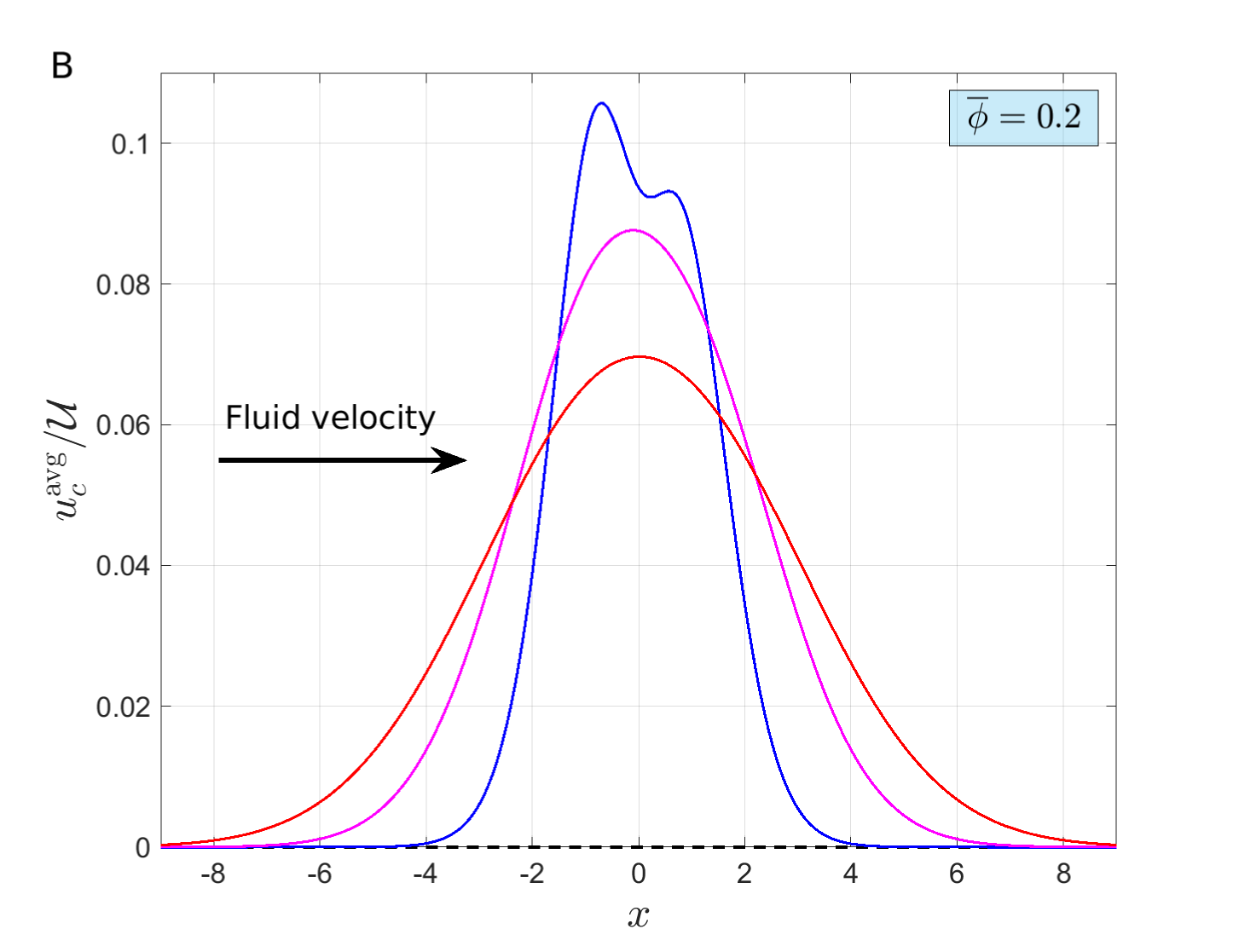}

\includegraphics[scale=0.31]{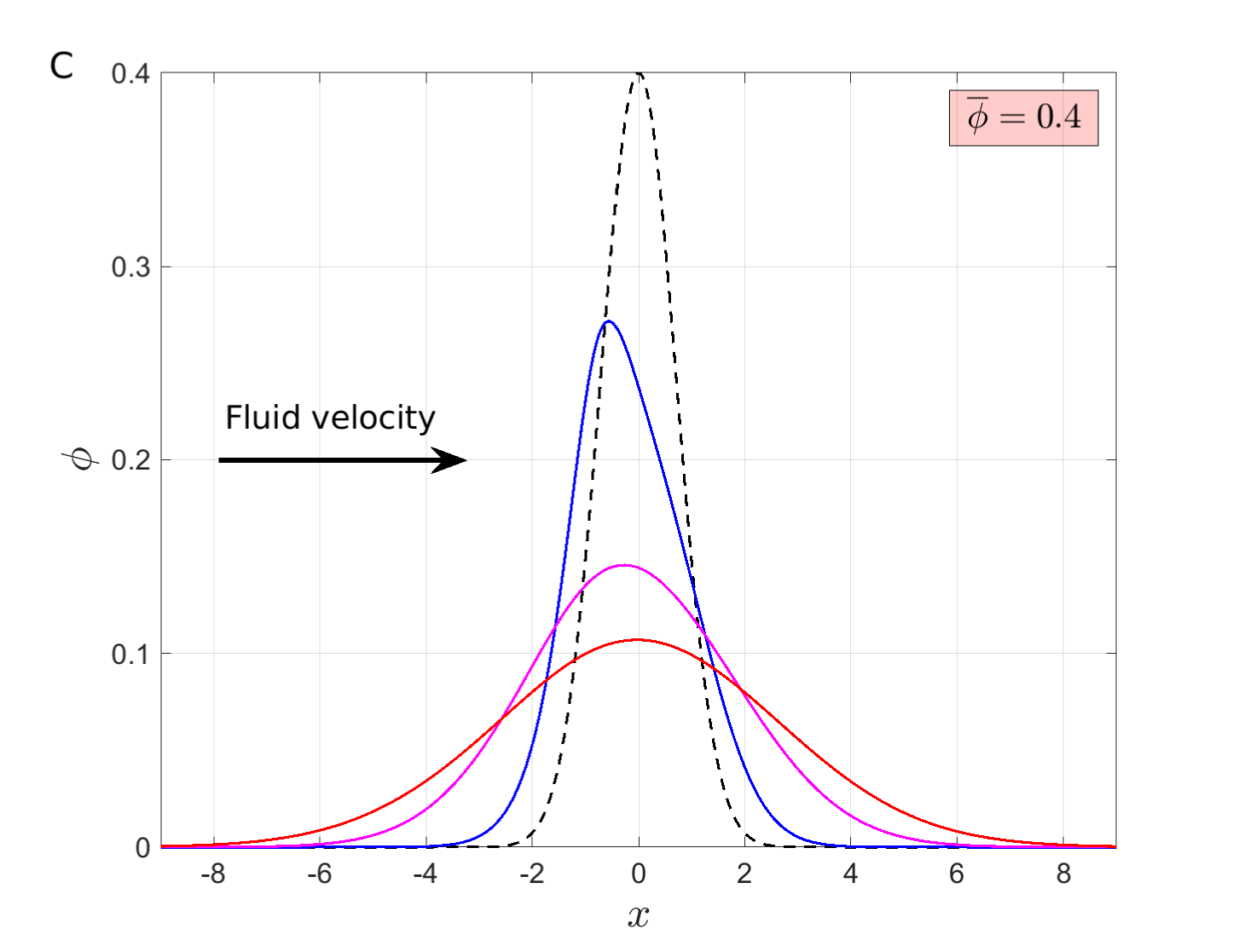}\hfill{}\includegraphics[scale=0.31]{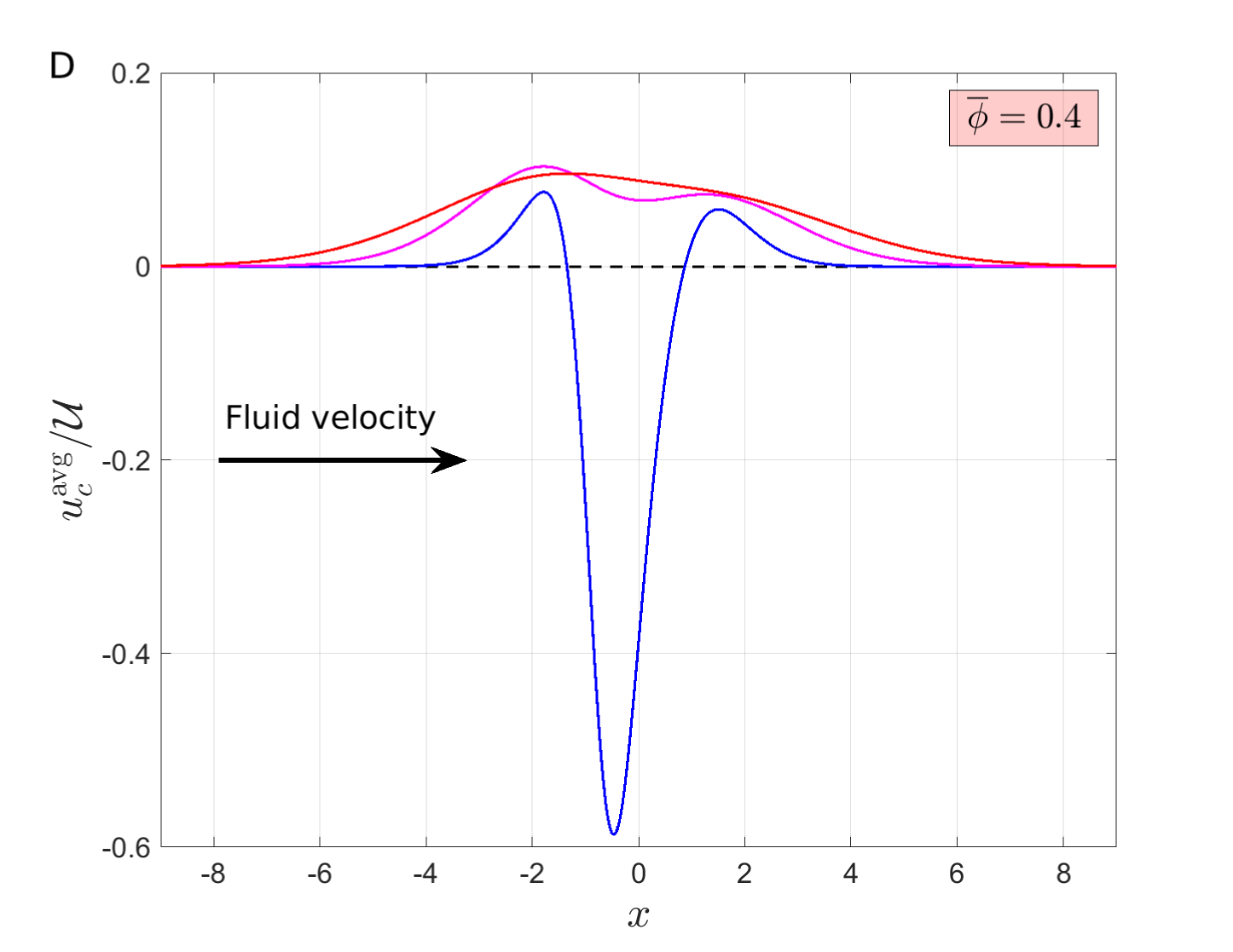}

\caption{The spatio-temporal evolution of the cell volume fraction (A,C) and
scaled average cell velocity (B,D) starting from a Gaussian $x$-distribution
of cells at $t=0$, with $\overline{\phi}=0.2$
(A,B) and $\overline{\phi}=0.4$ (C,D). The solutions are plotted at times $t=0$ (dashed-black line),
$t=100$ (blue), $t=500$ (magenta), and $t=1000$ (red). Parameter values: $\mathrm{Da}=0.5$, $\mathcal{T}=10$; other parameters are fixed at 
the values listed in Table~\ref{tab:Nondimensional-parameter-values}. The arrows indicate the direction of the fluid velocity (i.e., the downstream direction). Positive and negative velocities in (B,D) represent average cell velocities that are oriented downstream and upstream, respectively.
\label{fig:phi_and_u_c}}
\end{figure}

Figure~\ref{fig:phi_and_u_c} illustrates the spatio-temporal evolution
of the cell layer in response to the flow-induced chemotactic and
tensotactic stimuli. We define the macroscopic average cell velocity
as
\begin{equation}
u_{c}^{\mathrm{avg}}=\frac{\psi}{\phi},
\end{equation}
and plot the $x$-distributions of the scaled average velocity, $u_{c}^{\mathrm{avg}}/\mathcal{U}$
(macroscopic cell velocity normalized by the individual-cell velocity),
and the cell volume fraction, $\phi$, at times $t=0,100,500,1000$.
For the specific case of $\mathrm{Da}=0.5$ and $\mathcal{T}=10$,
we consider two initial cell volume fractions, $\overline{\phi}$, which
induce different migration behaviours: in Fig.~\ref{fig:phi_and_u_c}A and B,
$\overline{\phi}=0.2$, leading to dominant downstream migration at
all times, as can be seen from the average cell velocity profiles in Fig.~\ref{fig:phi_and_u_c}B which are always non-negative. By contrast, in Fig.~\ref{fig:phi_and_u_c}C and D, $\overline{\phi}=0.4$,
leading to upstream migration at early times, and a transition to
dominant downstream migration at later times (notice the dominant negative cell velocity at early times in Fig.~\ref{fig:phi_and_u_c}D, as indicated by the $t=100$ curve, which changes to an all-positive velocity profiles at $t>500$).

\begin{figure}[hbt!]
\begin{centering}
\includegraphics[scale=0.55]{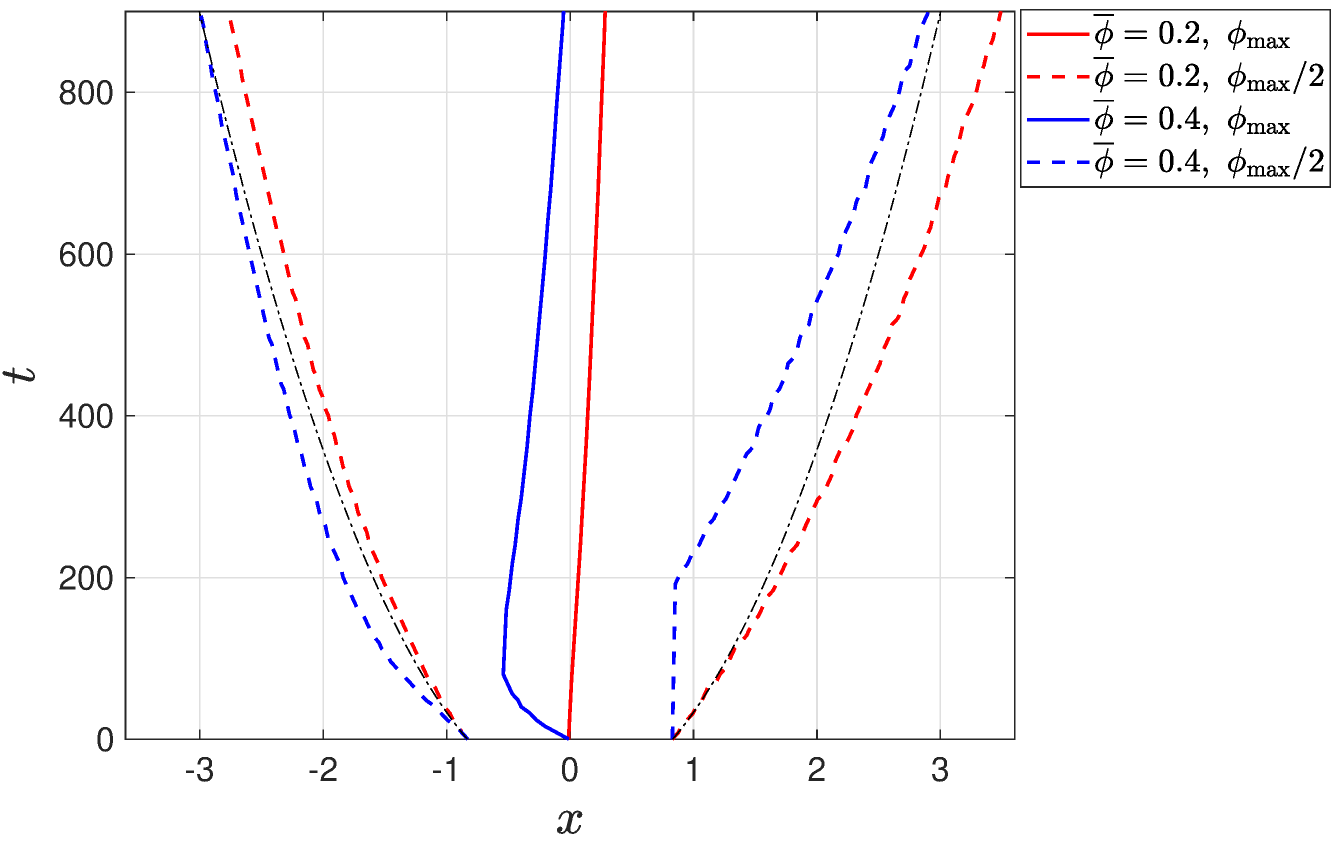}
\par\end{centering}
\caption{Different regions of the cell layer travel at different velocities. The change in the spatial position, $x$, of $\phi_{\mathrm{max}}$ (solid lines) and $\phi_{\mathrm{max}}/2$ (dashed lines) as a function of time, for the two cases presented in Fig.~\ref{fig:phi_and_u_c}, $\overline{\phi}=0.2$ (red lines) and $\overline{\phi}=0.4$ (blue lines). For comparison, the dashed-dotted black lines show the purely diffusive evolution of the spatial location of $\phi_{\mathrm{max}}/2$. \label{fig:x_t_const_phi}}
\end{figure}

To complete the picture of the different velocities in which different regions of the cell layer migrate, Fig.~\ref{fig:x_t_const_phi} shows how the spatial position, $x$, at which the volume fraction attains its maximal value, $\phi_{\mathrm{max}}$, changes over time $t$ (solid lines). Also shown are the trajectories of the two spatial locations at which the volume fraction attains its half-maximal value (dashed lines).
For purely diffusive (PD) motion (i.e., without directed migration, $\psi=0$), the cell flux is purely diffusive (i.e., in the direction of $-\partial \phi/ \partial x$.) 
The purely diffusive  trajectory, $x_{\mathrm{PD}}$, for which $\phi=\phi_{\mathrm{max}}/2$, is given by
\begin{equation}
x_{\mathrm{PD}}(t;\phi=\phi_{\mathrm{max}}/2)=\pm \ln({2})\sqrt{1+4t/\mathrm{Pe}_c}
\end{equation}
and is included in Fig.~\ref{fig:x_t_const_phi} for reference. Naturally, in the purely diffusive case the location of the maximal value does not change over time, $x_{\mathrm{PD}}(t;\phi=\phi_{\mathrm{max}})=0$.

Comparing the purely diffusive trajectories (black dashed-dotted lines) and the red trajectories ($\overline{\phi}=0.2$) in Fig.~\ref{fig:x_t_const_phi}, we note that both trajectories exhibit the same qualitative behaviour, except for a small, right (downstream) shift of the red curves due to the stimulus-induced directed motion. This means that when $\overline{\phi}=0.2$ we have a dominant cell diffusion, where the locations of  $\phi=\phi_{\mathrm{max}}/2$ predominantly travel in the direction opposite to the volume fraction gradient. This diffusive movement is superimposed on a small shift of the cell layer to the right (downstream) due to the chemotaxis-induced positive velocity. For low cell volume fractions the effect of the stimulus-induced migration is small due to the small average cell velocity (up to 10\% of the individual cell speed in the case of $\overline{\phi}=0.2$, see Fig.~\ref{fig:phi_and_u_c}B). This is because, although chemotaxis dominates tensotaxis,
the chemotactic signal is rather weak at low volume fractions. While the cell layer in Fig.~\ref{fig:phi_and_u_c}A seems to maintain its symmetry with respect
to the location of the maximum volume fraction, it is possible to detect a small
amount of symmetry breaking in the cell-velocity field due to the nonlinearity of
the tensotactic and chemotactic stimuli which attain their maximum
values at slightly different $x$-locations.

At a larger initial cell volume fraction, in Fig.~\ref{fig:phi_and_u_c}C,
we can notice that the cell layer is skewed to the left at early times
due to the large negative cell velocity at the centre of the cell
layer (solid blue line in Fig.~\ref{fig:x_t_const_phi}). This
is because the tensotactic signal is dominant in the region where the
volume fraction is large. At early times we also notice that the stimulus-directed velocity and diffusive velocity are opposite at the downstream part of the cell layer, while they reinforce each other, to generate a large upstream velocity, at the upstream part of the cell layer (compare the dashed-blue curves and purely-diffusive curves in Fig.~\ref{fig:x_t_const_phi}).
At later
times, we observe a reduction in the cell volume fraction due to the
action of cell diffusion, which consequently leads to a transition to dominant downstream migration (notice the change to positive velocity in Fig.~\ref{fig:phi_and_u_c}D and in the solid-blue line in Fig.~\ref{fig:x_t_const_phi} when $t \gtrsim 100$).

\begin{figure}[hbt!]
\begin{centering}
\includegraphics[scale=0.45]{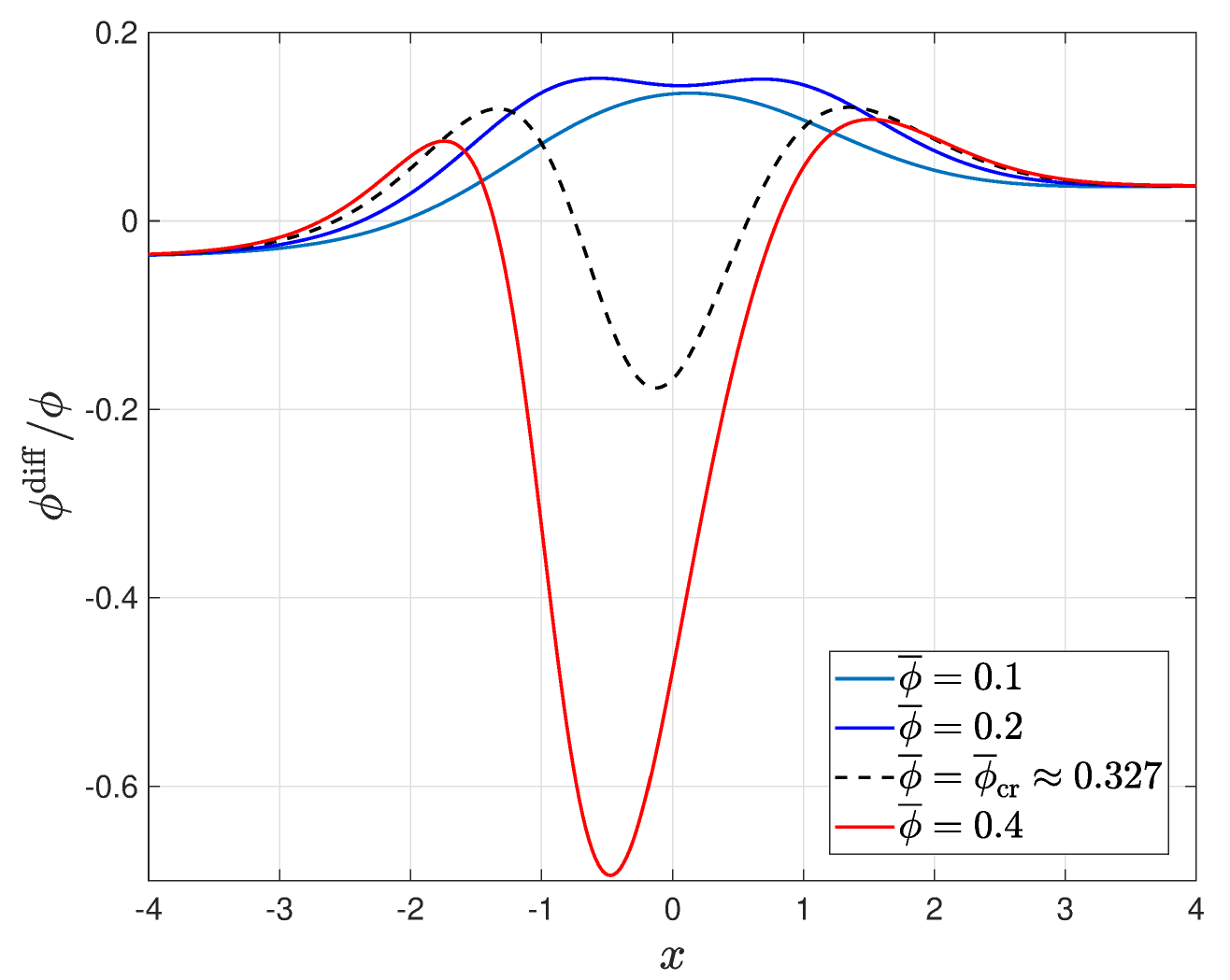}
\par\end{centering}
\caption{Series of plots showing how, at a fixed time point $t=100$, the proportion of cells moving upstream and downstream changes with $x$ for different values of the initial volume fraction, $\overline{\phi}$. The black dashed line represents the critical
volume fraction, $\overline{\phi}_{\mathrm{cr}}$, at which the dominant mode of migration switches  between downstream and upstream. Parameter values: $\mathrm{Da}=0.5$,
$\mathcal{T}=10$; other parameters use the values listed in Table~\ref{tab:Nondimensional-parameter-values}.\label{fig:phi_diff}}
\end{figure}

To better illustrate the change in the dominant mode of migration as the cell volume fraction changes, the results presented in Fig.~\ref{fig:phi_diff} show the spatial variation in the proportion of cells traveling downstream and upstream at time $t=100$ when $\mathrm{Da}=0.5$ and the initial volume fraction varies. For small $\overline{\phi}$, most cells travel downstream.
As $\overline{\phi}$ increases, more cells travel
downstream (compare the change between $\overline{\phi}=0.1$ and
0.2 in Fig.~\ref{fig:phi_diff}) due to the increased production of
chemokine; however, the maximum number of cells traveling downstream
is no longer in the centre of the cell layer due to the increased
tensotactic stimulus in the region where the cell volume fraction
is maximal. As $\overline{\phi}$ increases further, different migration directions dominate in different regions of the cell layer. On the one hand, the strong tensotactic stimulus in the centre of the cell layer, where the cell volume fraction is maximal, leads to upstream migration in this region; on the other hand, cells
in the edges, where the volume fraction is smaller, continue
to migrate downstream. At the critical value, $\overline{\phi}_{\mathrm{cr}}$
(dashed-black line in Fig.~\ref{fig:phi_diff}) there is a balance
between the proportion of upstream-migrating cells in the bulk of the
cell layer and the proportion of downstream-migrating cells at the 
edges of the cell cluster. As $\overline{\phi}$ increases beyond $\overline{\phi}_{\mathrm{cr}}$
more cells migrate upstream. Due to the strong nonlinearity of
the tensotactic stimulus, small deviations of $\overline{\phi}$ above
$\overline{\phi}_{\mathrm{cr}}$ amplify the tendency to upstream migration.
Consequently, the $x$-position where $\phi^{\mathrm{diff}}$
attains its minimal value (in the region of dominant tensotaxis) moves
to the left as $\overline{\phi}$ increases, because of the large
cell flux in the negative $x$ direction which shifts the location
of the maximum value of $\phi$.

Sufficiently far from the bulk of the cell layer, where the stimuli are
very weak and cell diffusion dominates, all curves of $\phi^{\mathrm{diff}}$
in Fig.~\ref{fig:phi_diff} collapse onto a single curve. In these
regions cells tend to diffuse in the direction of decreasing cell density.

\begin{figure}[hbt!]
\includegraphics[scale=0.32]{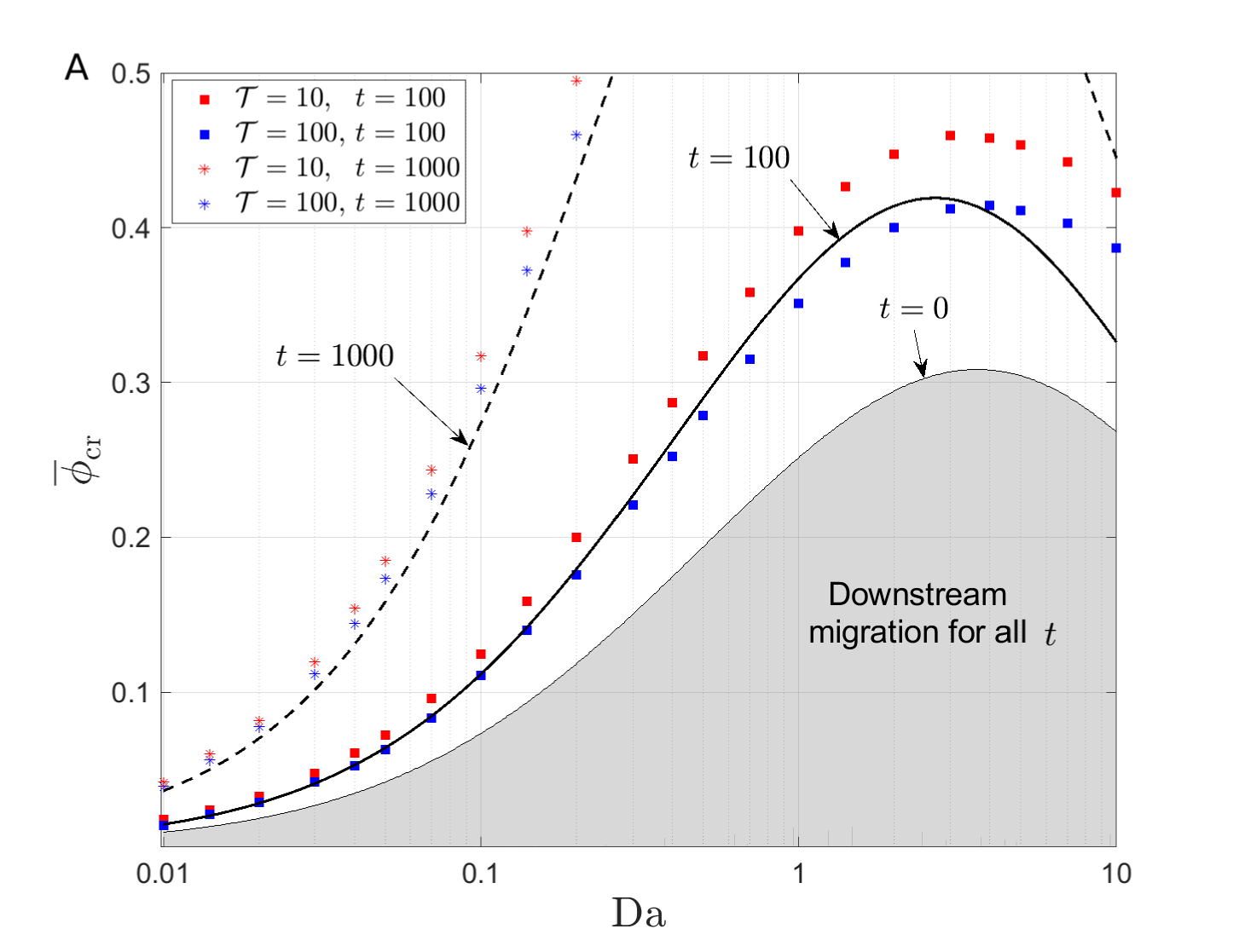}\hfill{}\includegraphics[scale=0.32]{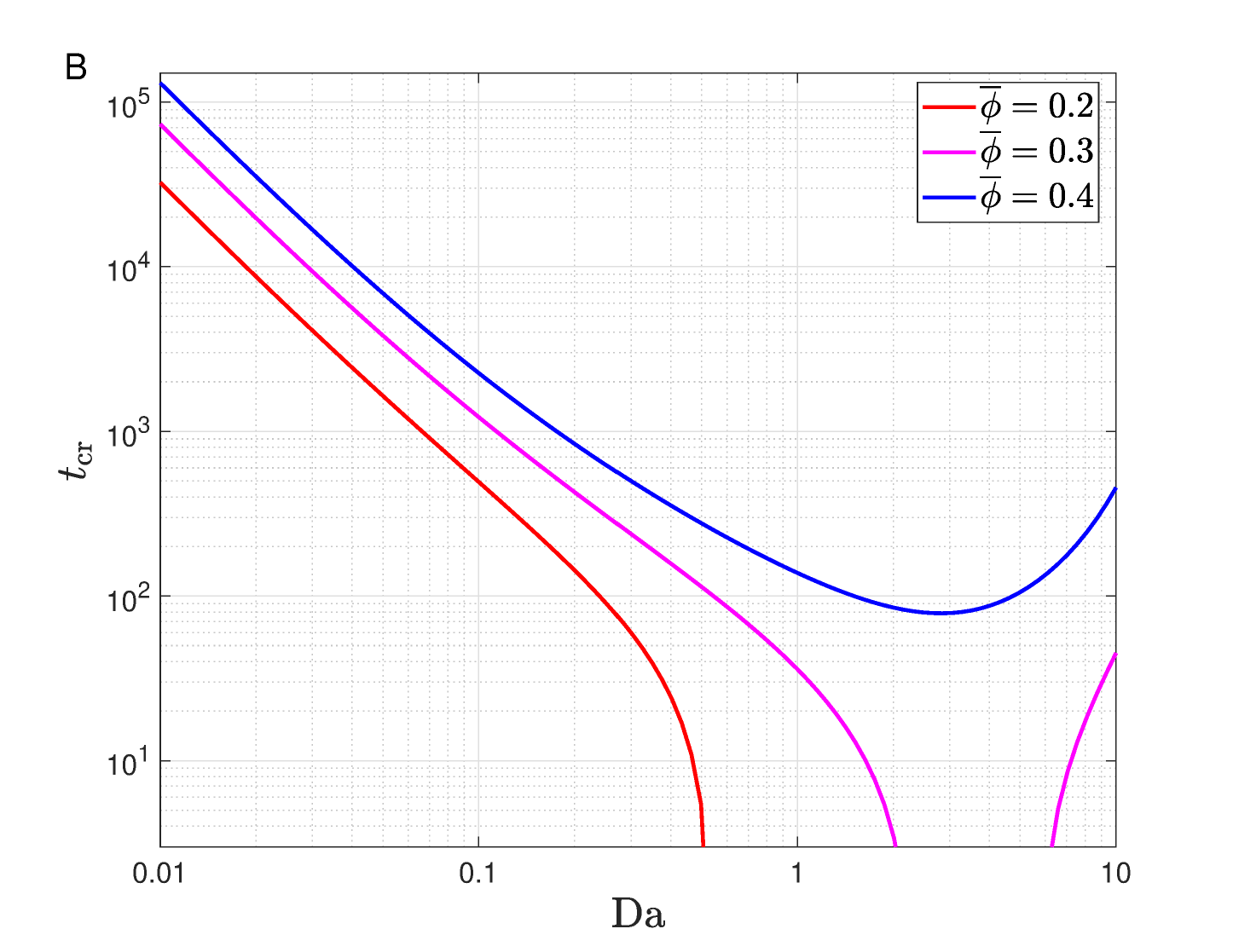}
\caption{The critical conditions for transition between dominant downstream and upstream migrations in the parameter space of $\mathrm{Da}$, $\overline{\phi}$, and $t$. (A) 
the critical cell volume fraction, $\overline{\phi}_{\mathrm{cr}}$, for transition as a function
of $\mathrm{Da}$. Comparison between the asymptotic expression [Eq.~(\ref{eq:phi_c_nonlinear})]
at times $t_{\mathrm{short}}=100$ (solid line) and $t_{\mathrm{long}}=1000$
(dashed line) and numerical simulation results at the respective times (square
and star symbols, respectively). Two values of cell relaxation
times, $\mathcal{T}=10$ (red symbols) and $\mathcal{T}=100$ (blue
symbols) were simulated. The shaded grey area indicates the region in parameter space in which upstream migration does not take place for any $t$. (B) the transition time, $t_{\mathrm{cr}}$, as a function
of $\mathrm{Da}$ for different values of $\overline{\phi}$, as predicted by the asymptotic model.
\label{fig:critical_phi}}
\end{figure}

Having observed transitions in the favourable migration direction as the system parameters vary, it is useful to delineate the parameter regimes in which these transitions occur. For that purpose, the results presented in Fig.~\ref{fig:critical_phi}A show how $\overline{\phi}_{\mathrm{cr}}$ changes as
the Damkohler number, $\mathrm{Da}$, is varied for two fixed values of the cell relaxation time, $\mathcal{T}$, and two values of the time that has elapsed since the initial condition; all other parameters are held fixed at their default values (see Table~\ref{tab:Nondimensional-parameter-values}). The characteristic
time scale is given by $L^{*}/U_{f}^{*}\sim100\,\mathrm{s}$; we used values
of $\mathcal{T}$ corresponding to a dimensional relaxation times of
several minutes ($\mathcal{T}=10$, $\tau^{*}\sim\,17\mathrm{min}$, red
symbols) and a few hours ($\mathcal{T}=100$, $\tau^{*}\sim3\,\mathrm{h}$, blue
symbols) corresponding to physiologically relevant relaxation times \citep{wyatt2016question}. For the time at which data were collected (i.e., the elapsed time since the start of the experiment), we used dimensional times of several hours
($t_{\mathrm{short}}=100$, square symbols) and approximately one day ($t_{\mathrm{long}}=1000$,
star symbols) to study cell behaviour on time scales which are either
much smaller or similar in magnitude to the timescale for cell migration, respectively. For each parameter combination we used the Matlab
function \texttt{fzero}, and our numerical scheme,  to determine the value of $\overline{\phi}$ for which
$N^{\mathrm{diff}}=0$ at the simulated time (either $t_{\mathrm{short}}$
or $t_{\mathrm{long}}$).

Figure~\ref{fig:critical_phi}A shows that $\overline{\phi}_{\mathrm{cr}}$
increases as the duration of time for which data are collected increases. This can be rationalised by noting that as $t$ increases the cell volume fraction decreases due to cell diffusion; thus, the tendency for upstream-oriented tensotactic migration decreases. Therefore, if the time at which we measure the mode of migration increases, then the initial volume fraction, $\overline{\phi}$, must also be increased, in order that a mode-transition (equal proportion of cells migrating down- and up-stream) occurs at this time. The grey region in Fig.~\ref{fig:critical_phi}A indicates the parameter region in which downstream migration prevails for all $t$. This region is delineated by the critical curve, $\overline{\phi}_{\mathrm{cr}}(\mathrm{Da};t=0)$, on which a ``transition'' between upstream and downstream migration occurs at time, $t=0$. Consequently, for each value of $\mathrm{Da}$, if $\overline{\phi}$ is outside this grey region, upstream migration will dominate at early times, and a transition to downstream migration will occur at some later time.
In more detail, a parameter combination in the area delineated by the two curves in Fig.~\ref{fig:critical_phi}A,  $\overline{\phi}_{\mathrm{cr}}(\mathrm{Da};t=t_1)$ and $\overline{\phi}_{\mathrm{cr}}(\mathrm{Da};t=t_2)$, will 
undergo a transition between upstream and downstream migration at some time in the interval $t\in(t_1,t_2)$.
Due to the action of cell diffusion, the cell volume fraction reduces over time and favours downstream migration at later times. Due to this mechanism, we expect
that at sufficiently large times downstream migration will always
prevail. However, depending on the values of the parameters, these times could be extremely long and may not be physiologically relevant. For example, the dimensional long time we used is equivalent to $t_{\mathrm{long}}^{*}\sim1\,\text{day}$.
This is about the maximum time scale on which the current model is
applicable, since at longer time scales, processes such as cell proliferation and death may no longer be negligible. To further illustrate the effect of the time that has elapsed since the beginning of the experiment on the migration behaviour, Fig.~\ref{fig:critical_phi}B shows the transition time, $t_{\mathrm{cr}}$, as a function of $\mathrm{Da}$ for a range of values of $\overline{\phi}$. In accordance with the grey region in Fig.~\ref{fig:critical_phi}A, for sufficiently small values of $\overline{\phi}$, there is a range of values of $\mathrm{Da}$ for which physically realistic values of $t_{\mathrm{cr}}$ do not exist (i.e., $t_{\mathrm{cr}} < 0$ in this region), and, thus, downstream migration prevails at all times. As expected, $t_{\mathrm{cr}}$ increases as $\overline{\phi}$ increases, reflecting the increase in tensotactic stimulus with the cell volume fraction increases.

While the time that has elapsed since the initial state may affect the value of $\overline{\phi}_{\mathrm{cr}}$
dramatically, Fig.~\ref{fig:critical_phi}A shows that varying
the cell relaxation time by a factor of 10 (while remaining in
the physiologically relevant regime) does not significantly alter the critical value of $\overline{\phi}_{\mathrm{cr}}$ (compare the red and blue symbols
in Fig.~\ref{fig:critical_phi}A). The modest increase in $\overline{\phi}_{\mathrm{cr}}$ for smaller relaxation times can be attributed to a more rapid reaction of the cells to changes in the dominant external stimulus from tensotaxis-dominated at early times to chemotaxis-dominated stimulus at later time. This, in turn, causes the transition to occur at earlier times.

The black lines in Fig.~\ref{fig:critical_phi}A correspond to the asymptotic
behaviour of $\overline{\phi}_{\mathrm{cr}}$ as $\mathrm{Da}\ll1$ and $\overline{\phi}\ll1$,
given by Eq.~(\ref{eq:phi_c_nonlinear}), for $t=100$ (solid line)
and $t=1000$ (dashed line). We note that the asymptotic model is
in excellent agreement with the numerical results when $\overline{\phi},\mathrm{Da}\ll1$. We note that it also replicates the general trend for larger values of $\overline{\phi}$
and $\mathrm{Da}$. As expected, the agreement improves when the cell
relaxation time increases or the elapsed time decreases. This
is because larger relaxation times and smaller elapsed times mean
less skewness of the cell distribution with respect to their $x$-symmetric
initial distribution, such that the assumptions 
on which
the asymptotic model is based [see Eq.~(\ref{eq:cell_limit_diffusion}) \textit{et
seq.}] are better fulfilled.

In the limit of $\overline{\phi}\ll1$ it can be readily seen from the asymptotic expression in Eq.~(\ref{eq:phi_c_nonlinear}) that
when $\mathcal{K}/\mathcal{M}$ decreases and $\mathrm{Da}$ increases
(for example, by reducing the fluid velocity), the critical volume
fraction for transition from downstream to upstream migration increases.
Indeed, in \cite{polacheck2011interstitial} smaller fluid velocities were found to reduce the tendency for upstream
migration (see Fig.~\ref{fig:exp_results}). Based on our model, we attribute this
behaviour to the impact that a reduction in the fluid velocity has on the mechanisms that inhibit upstream migration and
promote downstream migration: (i) the fluid-cell
drag force decreases, which leads to a smaller transcellular pressure
gradient and a smaller tensotactic cue; (ii) the ratio of reaction to advection increases, which leads to larger chemokine gradients and a larger downstream-oriented chemotactic signal. For large values of the Damkohler number ($\mathrm{Da}\gtrsim3$), we note a qualitative change in behaviour where further increases in $\mathrm{Da}$
lead to a reduction in $\overline{\phi}_{\mathrm{cr}}$ (i.e., a reduced tendency
to migrate downstream). This is due to increased 
chemokine consumption by the cells as the chemokine concentration increases,
which diminishes the chemotactic gradients in the downstream region
of the cell layer. This non-monotonic behaviour of the critical conditions with respect to the Damkohler number is also shown in Fig.~\ref{fig:critical_phi}B, where the transition time initially decreases with $\mathrm{Da}$ (corresponding to the aforementioned increase in the chemotactic stimulus), while starting to increase at $O(1)$ values of $\mathrm{Da}$.

\section{Conclusion}

The goal of the present study was to use mathematical modelling to study cell migration in response to flow-induced mechanical and chemical stimuli.
We developed a hybrid probabilistic-continuum model for a two-phase mixture of fluid and cells. We started from a mesoscopic-kinetic description for the cell probability density function, forced by a stimulus-dependent transition-probability function biasing the cell-velocity probability. Then, we used a velocity-space averaging to formulate a system of continuum equations that describe how the cells' spatial distribution evolves over time at the macroscopic level, in response to the mechanochemical signal. 
The use of a kinetic description as a starting point to derive continuum models for cell migration has been widely used [e.g., \cite{hillen2013transport,othmer2002diffusion,hillen2006m5,turner2004discrete,johnston2015modelling}]; additionally, there are several hybrid models of cell chemotaxis [e.g., \cite{dolak2005kinetic,filbet2014numerical,calvez2019chemotactic}], in which the transport of a chemoattractant is described by a macroscopic equation, with source/sink terms that depend on the density of cells, while the cells are described at the kinetic level. However, our model also accounts for tensotactic migration of cells. Since tensotactic migration depends on the pressure distribution, this necessitates the formulation of equations for fluid flow, which are coupled to cell motion via the incompressibility of the mixture and the dependence of the permeability on the cell volume fraction. To the best of our knowledge, coupling of this kind has not been carried out before.

Motivated by the experimental results by \cite{polacheck2011interstitial}, we focused on studying the migration of a one-dimensional cell layer in an infinite channel subject to a fluid flow. Contrary to purely continuum based models, our probabilistic approach enabled us to determine how the proportion of cells travelling upstream and downstream at a given spatial location evolves over time, and to determine the critical conditions at which transitions in the dominant mode of migration occur.

Through a combination of numerical simulation of the one-dimensional model and asymptotic analysis, we delineated the locus of transitions in the 
two parameter plane defined by the initial cell volume fraction, $\overline{\phi}$, and the Damkohler number, $\mathrm{Da}$, the latter parameter representing the ratio of chemokine secretion to advection rates.
In agreement with the experimental observations in \cite{polacheck2011interstitial} (see Fig.~\ref{fig:exp_results}), the current model predicts downstream-oriented chemotactic migration at low cell volume fractions, and upstream-oriented tensotactic migration at larger volume fractions. This effect can be understood by the increase in the transcellular pressure gradient and consequent tensotactic stimulus when the cell volume fraction increases.

In the experiments by \cite{polacheck2011interstitial}, the distribution of cell velocity was only measured at a single time point. By contrast, our model predicts that the time at which experimental data are collected has an important effect on the observed dominant mode of migration.
We identified a region of the parameter space in which the chemotactic stimulus dominates the tensotactic stimulus for all $t$ and, thus, downstream migration prevails for the duration of the experiment. By contrast, outside this region of parameter space, upstream migration prevails at the beginning of the experiment when the cells are localised, and a transition to downstream migration occurs at later times, due to the effect of cell diffusion, which causes the distribution of cells to become more dispersed over time. However, this initial dominant-upstream-migration transient can persist up to very long times ($t>1000 \rightarrow t^*>1\,\text{day}$, for some regions of parameter space). This means that it will be clearly visible in cell migration experiments [for example, in \cite{polacheck2011interstitial} measurements were taken after 24 hours]. This phenomenon may indicate the need to measure the cell velocities at different time points when conducting cell migration experiments.

We additionally showed that an increase in $\mathrm{Da}$ tends to increase the importance of chemotactic migration, due to enhanced chemokine secretion by the cells. However, our model predicts that there is an optimal value of $\mathrm{Da} \sim O(1)$ which maximizes the chemotactic signal; as $\mathrm{Da}$ increases above this local maximum, chemokine degradation increases, leading to smaller chemotactic gradients in the downstream region of the cell layer. Here we mention that the current results were obtained in the purely advective limit of the chemokine propagation, i.e., neglecting diffusive effects. It is expected that including chemokine diffusion and boundary interactions (e.g., no flux) will result in a more complicated behaviour.

Applying asymptotic analysis in the limit of $\phi,\mathrm{Da} \ll 1$ we obtained an explicit formula for the critical conditions in terms of the system parameters. The asymptotic expression showed an excellent agreement with the numerical results in the limit of $\phi,\mathrm{Da} \ll 1$, while it was also able to capture the general trend at larger values of $\phi$ and $\mathrm{Da}$, including the local maximum observed in the numerical results.

An important feature of the current model is the use of a permeability function that depends on the cell volume fraction, ensuring that tensotactic stimulus increases as the cell volume fraction increases. While in this work we used the isotropic Carman-Kozney permeability, in future work, it would be interesting to examine alternative permeability models that account for the anisotropic, fibrous nature of the ECM [e.g., \cite{happel1959viscous}].

While not considered in this paper, the kinetic model given by Eq.~(\ref{eq:f_BGK_2}) also allows one to calculate the microscopic cell-velocity probability distribution, $f(\mathbf{x},t,\boldsymbol{\xi})$. This would need to be carried out via a hybrid model where the numerical solution of $f$, obtained from the kinetic model in Eq.~(\ref{eq:f_BGK_2}), is averaged to produce the macroscopic cell volume fraction. This, in turn, is coupled to the fluid motion and chemoattractant transport, affecting the macroscopic stimulus profile; the stimulus is then fed back to the transition probability, $F$ in the kinetic equation. In future work, such a hybrid model could be used to evaluate the cell-velocity distribution, which could be compared to 
experimental measurements of the distribution of cellular velocities. In this way, it should be possible to refine the functional form of the transition-probability function, $F(\mathbf{x},t,\boldsymbol{\xi})$, to achieve better agreement with the experimental results.

Finally, the model developed in this paper constitutes a novel framework to study cell migration in a dynamic fluid environment. One example for such cell migration is the movement of tumour cells towards plasma-depleting blood vessels which can lead to either vessel collapse \citep{padera2004cancer} or intravasation \citep{roussos2011chemotaxis}. Here, the interaction of the cells with the vessel walls may affect the flux of interstitial fluid depleted by the vessel, thus coupling the extravascular cell migration to intravascular blood flow. This phenomenon has significant implications for tumour blood flow, progression, and therapy \citep{stylianopoulos2012causes,jain2014role}, and thus represents a natural topic for future work.

\section*{Acknowledgments}
This work was supported by the Engineering and Physical Sciences Research Council [grant number EP/X023869/1].

\section*{Data Availability}
The datasets generated during the current study are available from the corresponding author on reasonable request. The MATLAB code used to generate the data is available at the following GitHub repository: github.com/yaronbenami/cell\_migration.

\bibliography{cell_migration.bib}
\bibliographystyle{abbrvnat}

\end{document}